\documentclass[12pt]{article}
\usepackage{geometry}
\usepackage{amsmath}
\usepackage{amssymb}
\usepackage{amsmath}
\usepackage{booktabs}
\usepackage{graphicx}

\numberwithin{equation}{section}

\def\appendix#1{
  \addtocounter{section}{1}
  \setcounter{equation}{0}
  \renewcommand{\thesection}{\Alph{section}}
 \section*{Appendix \thesection\protect\indent \parbox[t]{11.715cm} {#1}}
  \addcontentsline{toc}{section}{Appendix \thesection\ \ \ #1}
  }

\renewcommand{\thefootnote}{\fnsymbol{footnote}}

\newcommand{\be}{\begin{equation}}
\newcommand{\ee}{\end{equation}}
\newcommand{\ba}{\begin{aligned}}
\newcommand{\ea}{\end{aligned}}

\newcommand{\hp}{\frac{p}{2}}

\def\m1{\left(-1\right)^{F_i}}

\makeatletter
\def\sla@#1#2#3#4#5{{%
  \setbox\z@\hbox{$\m@th#4#5$}%
  \setbox\tw@\hbox{$\m@th#4#1$}%
  \dimen4\wd\ifdim\wd\z@<\wd\tw@\tw@\else\z@\fi
  \dimen@\ht\tw@
  \advance\dimen@-\dp\tw@
  \advance\dimen@-\ht\z@
  \advance\dimen@\dp\z@
  \divide\dimen@\tw@
  \advance\dimen@-#3\ht\tw@
  \advance\dimen@-#3\dp\tw@
  \dimen@ii#2\wd\z@  \raise-\dimen@\hbox to\dimen4{%
    \hss\kern\dimen@ii\box\tw@\kern-\dimen@ii\hss}%
  \llap{\hbox to\dimen4{\hss\box\z@\hss}}}}
\def\slashed#1{%
  \expandafter\ifx\csname sla@\string#1\endcsname\relax
    {\mathpalette{\sla@/00}{#1}}%
  \else
    \csname sla@\string#1\endcsname
  \fi}
\makeatother

\def\ccw{{\hspace{-2.5mm}\unitlength 0.1in
\begin{picture}(1.00,1.00)(8.45,-11.50)
\special{pn 8}%
\special{pa 1120 1120}%
\special{pa 1100 1100}%
\special{pa 1080 1120}%
\special{fp}%
\end{picture}%
\hspace{-0mm}}}

\def\cw{{\hspace{-2.5mm}\unitlength 0.1in
\begin{picture}(1.00,1.00)(8.45,-11.50)
\special{pn 8}%
\special{pa 1120 1100}%
\special{pa 1100 1120}%
\special{pa 1080 1100}%
\special{fp}%
\end{picture}%
\hspace{-0mm}}}

\newcommand{\beq}{\begin{equation}}
\newcommand{\eeq}{\end{equation}}
\newcommand\beqa{\begin{eqnarray}}
\newcommand\eeqa{\end{eqnarray}}
\newcommand\bea{\begin{array}}
\newcommand\eea{\end{array}}

\newcommand{\nn}{\nonumber}
\newcommand{\neqa}{\nonumber\end{eqnarray}}
\newcommand{\la}{\label}
\newcommand{\J}{{\cal J}}

\newcommand{\OO}{{\cal O}}
\newcommand{\color}[1]{}

\newcommand{\eq}[1]{(\ref{#1})}

\newcommand{\h}{\hat}
\renewcommand{\t}{\tilde}

\def\({\left(}
\def\){\right)}
\def\[{\left[}
\def\]{\right]}

\def\<{\langle}
\def\>{\rangle}

\def\d{\partial}

\begin{document}

%%%%%%%%%%%%%%%%%%%%%%%%%%%%%%%%%%%%%
%%%%%%%%%%%%%%%%%%%%%%%%%%%%%%%%%%%%%
%%%%%%%%%%%%%%%%%%%%%%%%%%%%%%%%%%%%%

\thispagestyle{empty}
\begin{flushright}\footnotesize
%\texttt{arxiv:yymm.nnnn}\\
\texttt{CALT-68-2695}\\
%\texttt{LPTENS 08/NN}\\
%\texttt{SPhT-t08/NNN}\\
\vspace{2.1cm}
\end{flushright}

\renewcommand{\thefootnote}{\fnsymbol{footnote}}
\setcounter{footnote}{0}
\setcounter{figure}{0}
\begin{center}
{\Large\textbf{\mathversion{bold} Efficient precision quantization in AdS/CFT}\par}

\vspace{2.1cm}

\textrm{Nikolay Gromov$^{\alpha}$, Sakura Sch\"afer-Nameki$^{\beta}$ and Pedro Vieira$^{\gamma}$}
\vspace{1.2cm}

\textit{$^{\alpha}$ Service de Physique Th\'eorique,
CNRS-URA 2306 C.E.A.-Saclay, F-91191 Gif-sur-Yvette, France;
Laboratoire de Physique Th\'eorique de
l'Ecole Normale Sup\'erieure et l'Universit\'e Paris-VI,
Paris, 75231, France;
St.Petersburg INP, Gatchina, 188 300, St.Petersburg, Russia } \\
\texttt{nikgromov@gmail.com}
\vspace{3mm}

\textit{$^{\beta}$ California Institute of Technology\\
1200 E California Blvd., Pasadena, CA 91125, USA } \\
\texttt{ss299@theory.caltech.edu}
 \vspace{3mm}

\textit{$^{\gamma}$ Laboratoire de Physique Th\'eorique
de l'Ecole Normale Sup\'erieure et l'Universit\'e Paris-VI, Paris,
75231, France;  Departamento de F\'\i sica e Centro de F\'\i sica do
Porto Faculdade de Ci\^encias da Universidade do Porto Rua do Campo
Alegre, 687, \,4169-007 Porto, Portugal} \\
\texttt{pedrogvieira@gmail.com}
\vspace{3mm}

%%%%%%%%

\par\vspace{1cm}

\textbf{Abstract}\vspace{5mm}
\end{center}

\noindent
Understanding finite-size effects is one of the key open questions in solving planar AdS/CFT.
In this paper we discuss these effects in the $AdS_5 \times S^5$ string theory at one-loop in the world-sheet coupling.
First we provide a very general, efficient way to compute the fluctuation frequencies, which allows to determine the energy shift for very general multi-cut solutions. Then we apply this to two-cut solutions, in particular the giant magnon and determine the finite-size corrections at subleading order. The latter are then compared to the finite-size corrections from L\"uscher-Klassen-Melzer formulas and found to be in perfect agreement.

\vspace*{\fill}

\setcounter{page}{1}
\renewcommand{\thefootnote}{\arabic{footnote}}
\setcounter{footnote}{0}

\newpage

%%%%%%%%%%%%%%%%%%%%%%%%%%%%%%%%%%%%%
%%%%%%%%%%%%%%%%%%%%%%%%%%%%%%%%%%%%%
%%%%%%%%%%%%%%%%%%%%%%%%%%%%%%%%%%%%%

\tableofcontents
\newpage
%%%%%%%%%%%%%%%%%%%%%%%%%%%%%%%%%%%%%
%%%%%%%%%%%%%%%%%%%%%%%%%%%%%%%%%%%%%
%%%%%%%%%%%%%%%%%%%%%%%%%%%%%%%%%%%%%

\section{Introduction and Summary}

Semi-classical quantizing around generic classical configurations is a challenging problem in field theory.
In two-dimensional integrable field theories this situation is ameliorated, but it remains a difficult problem 
to quantize the theory around an arbitrary classical motion if we simply try to expand the action around the classical solution at stake.
In general  the quadractic Lagrangian will not be time independent, one needs to find the stability angles and the explicit determination of the fluctuation energy spectrum becomes computationally involved.
On the other hand, in general, classically integrable theories admit a finite gap description. 
In this construction each classical motion is mapped to a Riemann surface and semi-classical quantization amounts to 
pinching this surface by adding  extra singularities to the algebraic curve.  

The superstring in $AdS_5\times S^5$ falls precisely into this class of theories:
Sharpening our understanding of the quantum spectrum of the superstring in $AdS_5 \times S^5$ is of crucial importance.
However, despite much progress in semi-classical quantization of classical string configurations in $AdS_5 \times S^5$, 
it has remained a daunting problem to quantize around a generic classical string solution. 
Applying the conventional methods of semi-classical quantization becomes particularly challenging for so-called multi-cut solutions, which in terms of the finite-gap description correspond to higher-genus curves. 

The classical $AdS_5 \times S^5$ world-sheet theory however precisely admits a finite-gap description in terms on an algebraic curve
\cite{Kazakov:2004qf, Kazakov:2004nh, Beisert:2004ag, SchaferNameki:2004ik, Beisert:2005bm}.  Each classical string motion maps to a Riemann surface, and semi-classical quantization can be performed by pinching this curve \cite{Gromov:2007aq}. This approach has been 
successfully applied to various string configurations, and were shown to reproduce 
the standard world-sheet results of \cite{Frolov:2002av, Frolov:2003tu, Park:2005ji}.

In this paper we propose a very general, efficient quantization method, 
which is applicable to a very large class of classical string configurations. We will derive this within the framework of the algebraic curve. 
The key idea that it is based on is the concept of off-shell fluctuation energies, which we advocate in the main text, 
 and allows one to find the full spectrum around a vast set of classical solutions from the knowledge of 
\beq
\text{one $S^3$ and one $AdS_3$ fluctuation frequency ("frequency basis").}
\eeq
Compared to standard semi-classical quantization of string solutions one is not required to compute the fluctuations in all bosonic and fermionic fields. In particular it is interesting to note that the fermionic excitations can be constructed from these building blocks alone. Furthermore, this method is applicable to multi-cut solutions, which are from the conventional point of view, difficult to quantize. The concept of quasi-energy for the $\mathfrak{su}(2)$ sector, which is related to that of off-shell frequency, was introduced earlier in \cite{Vicedo:2008jy}. We will show however, that this is not only an abstract concept but can be put to practical use.

We will demonstrate the efficiency of our quantization method by computing the semi-classical spectrum 
 around the (dyonic) giant-magnon solution \cite{Hofman:2006xt, Dorey:2006dq, Chen:2006gea}. This describes a  classical string
moving in $S^3$ with angular momenta $J$ and $Q$ and world-sheet momentum $p$.  
When $J\to \infty$ this solution becomes the fundamental excitations of the two dimensional field theory
 defined in the infinite volume and its dispersion relation reads
\beq
\epsilon_\infty(p)=\sqrt{Q^2 + {\lambda\over \pi^2} \sin^2 {p\over 2}} \,.
\eeq
When $J$ is large but not infinite this expression receives exponential corrections \cite{Arutyunov:2006gs,Astolfi:2007uz,Minahan:2008re} which can be physically traced back to the existence of wrapping interactions \cite{Luscher:1985dn,Klassen:1990ub,Ambjorn:2005wa,Janik:2007wt,Hatsuda:2008gd,Hatsuda:2008na,Gromov:2008ie,Heller:2008at,Bajnok:2008bm}. 

The giant magnon solution is paradigmatic for the efficiency of this approach. For example, 
in infinite volume $\epsilon_\infty(p)$ has no constant term in the large $\lambda$ expansion. 
This means that the one-loop shift should vanish. 
Showing this fact from a direct world-sheet field theory computation is a rather involved computation \cite{Papathanasiou:2007gd}, whereas 
from the point of view of the algebraic curve point  this result is obtained in a trivial way \cite{Chen:2007vs,Gromov:2008ie}. 
%Let us review the argument.
%When we add a quantum fluctuation as in figure \ref{CurveSheets} the fluctuation will have some energy $\epsilon(x)$ where $x$ is the position where the fluctuation is located in the curve and some momentum $p(x)$. The quantum fluctuation energy is given by the energy shift due to $\epsilon(x)$ plus the backreaction which is clearly $\epsilon_\infty'(p)\delta p$ where $\delta p$, the shift in momentum of the classical magnon must be equal to $-p(x)$ by momenta conservation. Thus, for any fluctuation in figure \ref{CurveSheets} we find a fluctuation energy $E_n^{ij}=\Omega^{(0)}(x_n^{ij})$. From out discussion That, is the function $\Omega^{(0)}$

%%%%%%%%%%%%%%%%%%%%%%%%%%%%%%%
\begin{figure}
\begin{center}
  \includegraphics*[width =12cm]{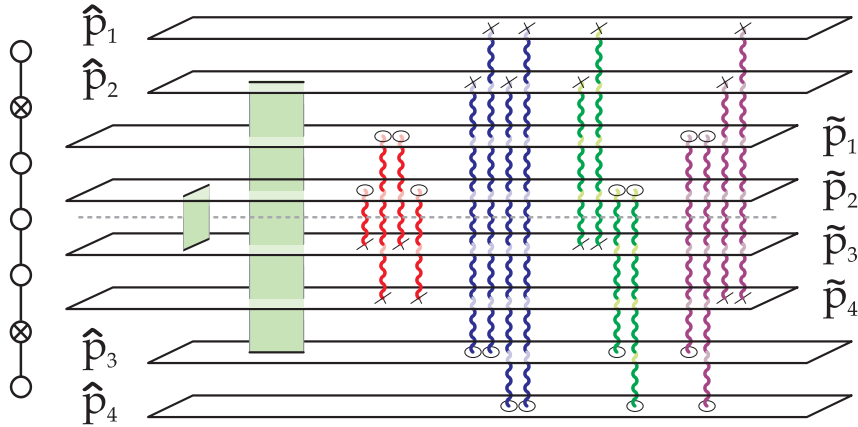}  \caption{\small Algebraic curve for classical superstrings on $AdS_5 \times S^5$. The macroscopic green cuts corresponds to a classical configuration. The wavy lines depict the several physical fluctuations. From left to right we have four bosonic $S^5$ fluctuations , four $AdS_5$ and eight fermionic fluctuations respectively. Any physical configuration has to cross the dashed line. We depict only the physical $|x|>1$ region. \label{CurveSheets}
 }
  \end{center}
\end{figure}
%%%%%%%%%%%%%%%%%%%%%%%%%%%%%%%%%%%

For the quantization of the finite volume dyonic giant magnon a direct world-sheet approach would most certainly be incomparably more involved than the one based on the classical algebraic curve which we carry out in this paper.
Needless to say the method we propose here and illustrate with the giant magnon solution can be applied to other integrable theories and to very different classes of classical solutions.

The plan of the paper is as follows: 
we will begin with a lightning review of the algebraic curve and its semi-classical quantization. In section \ref{sec:QTools} we prove our efficient quantization method and provide a closed formula for the one-loop energy shift in terms of our "frequency basis".
In section \ref{sec:GeneralTwoCut} this approach is exemplified for the generic two-cut $\mathfrak{su}(2)$ solution and in section
\ref{sec:GMTwoCut} we compute the energy shift to the giant magnon and extract the subleading correction. Finally in section \ref{sec:Luscher} we compute these corrections from the L\"uscher-Klassen-Melzer formulas and show their agreement with our semi-classical quantization method. 

Throughout all the paper we use
\beq
g=\frac{\sqrt{\lambda}}{4\pi} \,,\qquad 
\mathcal{E}=\frac{\Delta}{\sqrt{\lambda}} \,,\qquad 
 \J=\frac{J}{\sqrt{\lambda}} \,,\qquad 
  \mathcal{Q}=\frac{Q}{\sqrt{\lambda}}\,.
\eeq

%%%%%%%%%%%%%%%%%%%%%%%%%%%%%%%%%%%%%
%%%%%%%%%%%%%%%%%%%%%%%%%%%%%%%%%%%%%
%%%%%%%%%%%%%%%%%%%%%%%%%%%%%%%%%%%%%

\section{Quantizing the algebraic curve}
\label{sec:SemiClassics}

%%%%%%%%%%%%%%%%%%%%%%%%%%%
%%%%%%%%%%%%%%%%%%%%%%%%%%%

\subsection{Classical algebraic curve}

In \cite{Kazakov:2004qf} a beautiful map between classical superstring motion in $AdS_5\times S^5$ and Riemann surfaces was presented. The idea is that using the Bena-Polchinski-Roiban \cite{Bena:2003wd} flat connection $A(x)$ -- where $x$ is an arbitrary complex number, the so-called spectral parameter -- we can diagonalize the monodromy matrix
\beq
\Omega(x)={\rm Pexp\,}\oint_\gamma A(x) \label{mono} \,,
\eeq
where $\gamma$ is any path starting and ending at some point $(\sigma,\tau)$ and wrapping the worldsheet cylinder once, to obtain a set of (eight) eigenvalues
\beq
\{e^{i\h p_1},e^{i\h p_2},e^{i\h p_3},e^{i\h p_4}|e^{i\t p_1},e^{i\t p_2},e^{i\t p_3},e^{i\t p_4} \} \,,
\eeq
which, due to flatness of the current, are $\gamma$ independent. As they depend on the arbitrary complex number $x$ they give rise to conserved charges by Taylor expansion around any point in the $x$-plane.
Since they are obtained from the diagonalization of an (almost) regular matrix $\Omega(x)$ they are obtained by solving a characteristic equation and thus define an (eight-sheeted) algebraic curve. The properties of this curve \cite{Kazakov:2004qf} follow from those of the flat connection $A(x)$ and are summarized
in appendix A.

The quasimomenta $p_i(x)$, being the log of the eigenvalues of $\Omega(x)$, do not define a Riemann surface. Rather, when evaluated on the algebraic curve for $e^{i p_i(x)}$ they might jump by an integer multiple of $2\pi$ as one crosses one of the square root cuts of the algebraic curve, i.e.
\beq
p_i^+ (x)-  p_j^-(x)=2\pi n_{ij} \,\, , \,\, x\in \mathcal{C}_{n}^{ij} \label{AlgCur} \,,
\eeq
where $p^{\pm}_i(x)$ is the value of the quasimomentum above/bolow the cut.
Moreover to each cut we can associate a filling fraction given by integrating the quasimomenta around the cut
\beq\label{FillFrac}
S_{ij}=\pm\,\frac{\sqrt{\lambda}}{8\pi^2i}\oint_{\mathcal{C}_{ij}} \(1-\frac{1}{x^2}\) p_i(x) dx \,.
\eeq
Thus, each cut of the algebraic curve is characterized by a discrete label $(i,j)$, corresponding to the two sheets being united, an integer $n$, the multiple of $2\pi$ mentioned above, and a real filling fraction. These three quantities are the analogues of the polarization, mode number and amplitude of the flat space Fourier decomposition of a given classical solution. The (sixteen) superstring physical polarizations correspond to the pairing of sheets
\beq\label{Polarisations}
\begin{aligned}
  S^5 :& \quad (\t 1,\t 3)\,,(\t 1,\t 4)\,,(\t 2,\t 3)\,,(\t 2,\t 4) \cr
AdS_5 :& \quad (\h 1,\h 3)\,,(\h 1,\h 4)\,,(\h 2,\h 3)\,,(\h 2,\h 4) \cr
\text{Fermions}
      :& \quad (\t 1,\h 3)\,,(\t 1,\h 4)\,,(\t 2,\h 3)\,,(\t 2,\h 4)\cr
       & \quad (\h 1,\t 3)\,,(\h 1,\t 4)\,,(\h 2,\t 3)\,,(\h 2,\t 4) \,.
\end{aligned}
\eeq
These physical polarizations are determined by the constraint that the lines connecting the sheets $(ij)$ have to cross the yellow line in figure \ref{CurveSheets}.
A simple rule of thumb is that they always connect sheets with index $1$ or $2$ with $3$ or $4$.
The classical energy of the string is obtained from the asymptotics (\ref{AsympInfty}) 
\beq
E  = {\sqrt{\lambda} \over  4 \pi} \lim_{x\to\infty} x \left(\hat{p}_1 (x) + \hat{p}_2 (x)\right) \,.
\eeq
%In Appendix $C$ we give an explicit example of such an algebraic curve construction.

%%%%%%%%%%%%%%%%%%%%%%%%%%%%%%

\subsection{Quantization}

Semi-classical quantisation of the algebraic curve proceeds by adding small number of fluctuations on top of the classical configuration \cite{Gromov:2007aq}. This treatment is equivalent to the semi-classical computation of quadratic fluctuations in the sigma-model \cite{Frolov:2002av, Frolov:2003tu, Park:2005ji}, however we will show that the algebraic curve approach is far more efficient.

We consider fluctuations around the classical curve for each polarization $(i, j)$ and mode number $n$. Adding a fluctuation amounts to shifting the quasimomenta as
$p_k (x) \rightarrow p_k (x) + \delta^{ij}_n p_k (x)$ where $\delta^{ij}_n p_k(x)$ is constrained by precise analytical properties as listed in Appendix A.2. In particular, the quasimomenta $\delta^{ij}_n p_i$ and $\delta^{ij}_n p_j$ which are the quasimomenta connected by the fluctuation at stake must behave as
\beq\label{PolePosition}
\delta^{ij}_n p_i (x) \simeq \pm \frac{\alpha(x_n^{ij})}{x-x_n^{ij}}
\eeq
close to  the pole position $x_n^{ij}$ which is determined by
\beq
p_i (x_n^{ij}) - p_j (x_n^{ij}) = 2 \pi n_{ij} \,. \label{PolePosition}
\eeq
The physical poles correspond to solutions of this equation with $|x_n^{ij}|>1$. The precise choice of signs above as well as $\alpha(y)$ is given in Appendix A.2. Having found $\delta^{ij}_n p_k$ we read off the fluctuation energy with mode number $n$ and polarization $(i,j)$ from the large $x$ asymptotics
\beq
\Omega_n^{ij}  =-2\,\delta_{i,\h1}+ {\sqrt{\lambda} \over  2 \pi} \lim_{x\to\infty} x \, \delta_n^{ij}\hat{p}_1 (x)  \,. \la{largex}
\eeq
In the next section we will explain that in fact we do not need to compute separately each of the sixteen physical fluctuations corresponding to the various string polarizations (\ref{Polarisations}) but that it suffices to compute two of them, at least for a huge number of interesting solutions. In particular we shall see that the fermionic fluctuations can be obtained from the $S^3$ and $AdS_3$ fluctuation energies.

%%%%%%%%%%%%%%%%%%%%%%%%%%%%%%%%%%%%%%%%%
%%%%%%%%%%%%%%%%%%%%%%%%%%%%%%%%%%%%%%%%%

\subsection{Quantizers toolkit}
\label{sec:QTools}

Notice that the dependence on $n$ of the shift in the quasimomenta $\delta_n^{ij} p_k$ only appears through $x_n^{ij}$ as determined in (\ref{PolePosition}). In other words the shift in the quasimomenta is actually a function of the position of the pole, i.e.
\beq
\delta_n^{ij} p_k(x)=\left. \delta^{ij} p_k(x;y) \right|_{y=x_n^{ij}} \,.
\eeq
Moreover the \textit{off-shell} quantity $ \delta^{ij} p_k(x;y)$ is a well defined function of $y$. It is determined by the same asymptotics as for the \textit{on-shell} shift of quasimomenta $\delta_n^{ij} p_k(x)$ except that the position of the pole is left unfixed.  An obvious consequence of what we just observed is that the fluctuation energies read off from (\ref{largex}) are, by construction, of the form
\beq
\Omega_n^{ij}=\left.\Omega^{ij}(y)\right|_{y=x_n^{ij}} \,,
\eeq
where the function $\Omega^{ij}(y)$ is independent of the mode number $n$. We call $\Omega^{ij}(y)$ the \textit{off-shell} fluctuation energies. The off-shell frequency is related for the particular case of the $SU(2)$ principal chiral model to the quasi-energy introduced in  \cite{Vicedo:2008jy}.

Given an on-shell fluctuation energy $\Omega_n^{ij}$ as a function of the mode number $n$, we can always reconstruct the off-shell frequencies by first computing the quasimomenta $p_i(x)$ for the underlying classical solution and then we simply replace $n$ using (\ref{PolePosition}), that is
\beq
\Omega^{ij}(y) =\left.  \Omega_n^{ij} \right|_{n\to \frac{p_i(y)-p_j(y)}{2\pi}} \,.
\eeq
In Appendix C this is exemplified for a simple $S^3$ circular string.

We will now explain how, using the inversion symmetry (\ref{Auto}), we can relate the several off-shell fluctuation energies. In this way we will find a powerful reduction algorithm for the computation of the fluctuation energies and thus the one loop energy shift
\beq\label{OneLoopE}
\delta \Delta^{1-loop}=\frac{1}{2} \sum_{ij,n}(-1)^{F_{ij}} \Omega_n^{ij} \,,
\eeq
around a generic classical solution.

%%%%%%%%%%%%%%%%%%%%%%%%%%%%%%%

\subsubsection{Frequencies from inversion symmetry}

\begin{figure}
\begin{center}
  \includegraphics*[width =15cm]{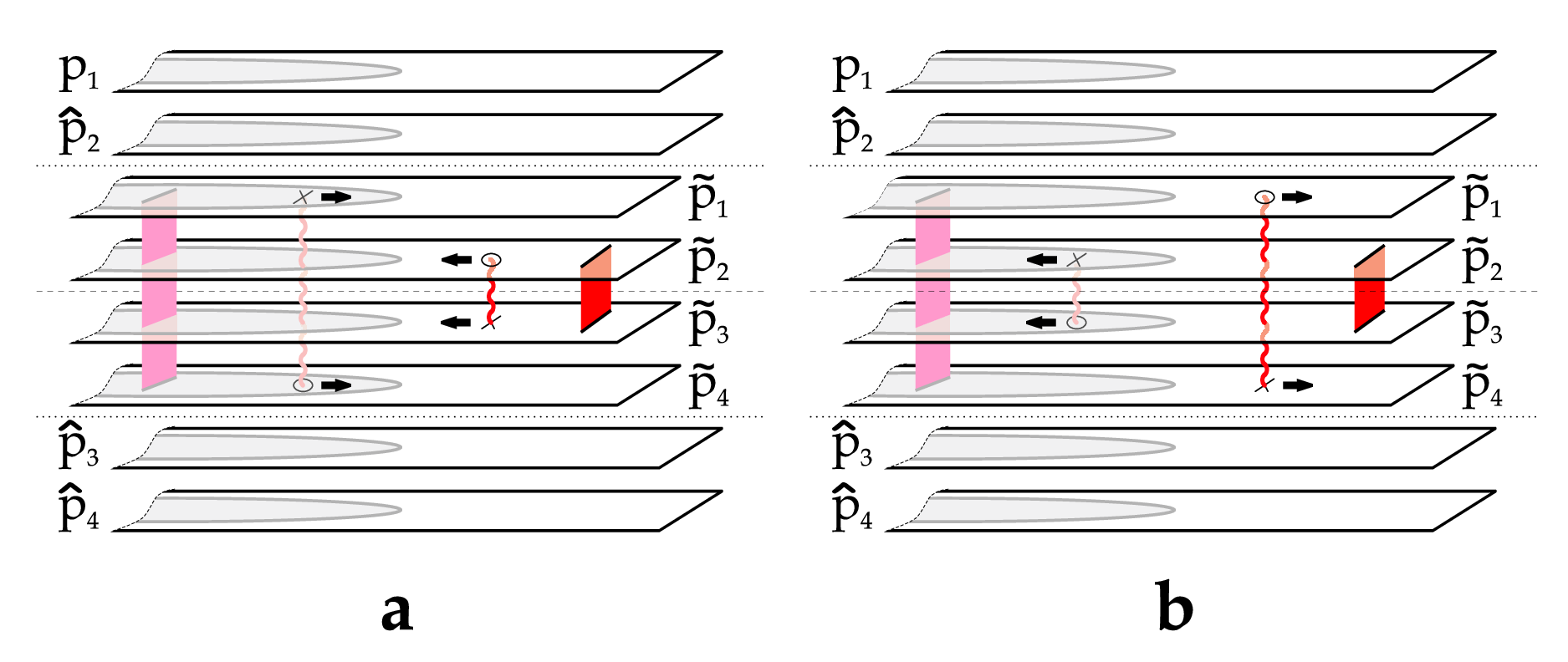}  \caption{\small As we analytically continue a fluctuation energy $\Omega^{\t2 \t3}(y)$ from a point $|y|>1$ to the interior of the unit circle we see that its mirror image becomes physical. \label{Exchange}
 }
  \end{center}
\end{figure}

An important property of the quasi-momenta, which follows from the $\mathbb{Z}_4$-grading of the $\mathfrak{psu}(2,2|4)$ superalgebra, is the inversion symmetry (\ref{Auto}) under $x\rightarrow 1/x$, which exchanges the quasi-momenta $p_{\t1,\t4} \leftrightarrow p_{\t2,\t 3}$ and likewise for the $AdS$ hatted quasi-momenta.  Thereby, a  pole connecting the sheets $(\t2, \t3)$ at position $y$, always comes with an image pole at position $1/y$ connecting the sheets $(\t1,\t4)$.
We can obtain a physical frequency  $\Omega^{\t 1\t 4}(y)$, by analytically continuing the off-shell frequency $\Omega^{\t 2 \t 3}(y)$, inside the unit circle. This is because when we cross the unit-circle, the physical pole for $(\t2\t3)$ becomes unphysical, thereby rendering its image, which lies now outside the unit-circle, a physical pole for $(\t1\t4)$ as depicted in figure \ref{Exchange}.

Let us consider in detail how this works for the $AdS$ fluctuations. As we will now demonstrate
\beq\label{OmegaMapAdS}
\Omega^{\h1\h4}(y)=-\Omega^{\h2 \h3}(1/y)-2 \,.
\eeq
Thus, suppose we know $\Omega^{\h 2 \h 3}(y)$. We know that this fluctuation energy appears  in the asymptotics of the shifted quasimomenta $\delta^{\h 2 \h 3} p_k(x;y)$ defined by the analytic properties listed in Appendix A.2.  Consider now $ - \delta^{\h 2 \h 3} p_k(x;1/y)$. From the analytic properties of $ \delta^{\h 2 \h 3} p_k(x;y)$ we conclude that
\begin{itemize}
\item
Close to $x=y$ we have
\beq
- \delta^{\h 2 \h 3} p_{\h 1}(x;1/y) \simeq \frac{\alpha(y)}{x-y} \,,\qquad 
 - \delta^{\h 2 \h 3} p_{\h 4}(x;1/y) \simeq - \frac{\alpha(y)}{x-y} \,.
\eeq
 \item
The poles at $x=\pm 1$ for these functions $- \delta^{\h 2 \h 3} p_k(x;1/y)$ are also synchronized as in equation (\ref{DeltapVirConst}).
\item
Close to the branch points of the original solution these functions exhibit inverse square root singularities.
\end{itemize}
These are precisely the required properties for  $\delta^{\h 1 \h 4} p_k(x;y)$ as listed in Appendix A.2! Thus
\beq
\delta^{\h 1 \h 4} p_k(x;y)  =- \delta^{\h 2 \h 3} p_{ k}(x;1/y) \la{ident} \,.
\eeq
From the large $x$ asymptotics we have
\beqa
- {\sqrt{\lambda} \over  4 \pi} \lim_{x\to\infty} x\, \delta^{\h2\h3}\hat{p}_{\h 1} (x;1/y) &=&- \frac{\Omega^{\h2\h3}(1/y)}{2}\,,
%\\ -{\sqrt{\lambda} \over  4 \pi} \lim_{x\to\infty} x\, \delta^{\h2\h3}\hat{p}_{\h 2} (x;1/y) &=& -\frac{\Omega^{\h2\h3}(1/y)}{2}-1
\eeqa
while by definition $\Omega^{\h 1 \h 4}(y)$ can be read off from
\beqa
 {\sqrt{\lambda} \over  4 \pi} \lim_{x\to\infty} x\, \delta^{\h1\h4}\hat{p}_{\h 1} (x;y) &=& \frac{\Omega^{\h1\h4}(y)}{2}+1 \,.
 %\\ {\sqrt{\lambda} \over  4 \pi} \lim_{x\to\infty} x\, \delta^{\h1\h4}\hat{p}_{\h 2} (x;y) &=& \frac{\Omega^{\h1\h4}(y)}{2}
\eeqa
From the identification
(\ref{ident})
we thus conclude (\ref{OmegaMapAdS}).

Similarly we can proceed for the $S^5$ frequencies and relate $\Omega^{\t2\t3}(y)$ with $\Omega^{\t1 \t4}(y)$.
It is clear that $\Omega^{\t1\t4}(y) = -  \Omega^{\t2 \t3}(1/y)$ +constant, and to find this constant we can either repeat the analysis we just did applied to the sphere fluctuations or we can be smarter and fix it from $\Omega^{\t1\t4}(\infty)=0$. This must of course hold -- the energy shift when we add an extra root at infinity is obviously zero, in other words, roots at infinity are  zero modes.
Thus,  the relation we find is similar to (\ref{OmegaMapAdS}), except that the constant term differs:
\beq\label{InversionOmega}
\Omega^{\t1\t4}(y) = -  \Omega^{\t2 \t3}(1/y) +\Omega^{\t2 \t3}(0)\,.
\eeq
Obviously for the purpose of computing the one-loop shift these constants are irrelevant as they will cancel in the sum.

So far we have obtained the frequencies $(14)$ from $(23)$. In the next subsection we will show how to derive all remaining frequencies. For a very large class of classical solutions we will be able to extract all fluctuation energies, including the fermionic ones, from the knowledge of a single $S^3$ and a single $AdS_3$ fluctuation energy.

%%%%%%%%%%%%

\subsubsection{Basis of fluctuation energies}

For simplicity let us consider only symmetric classical configurations that have pairwise symmetric quasi-momenta
\beq\label{SymSol}
p_{\h1, \h 2 ,\t 1, \t2 } = - p_{\h4, \h3, \t 4 , \t3} \,,
\eeq
as depicted in figure \ref{CurveSheets}.
This is in particular the case for all rank one solutions, i.e. $\mathfrak{su}(2)$ and $\mathfrak{sl}(2)$.

\begin{figure}
\begin{center}
  \includegraphics*[width =16cm]{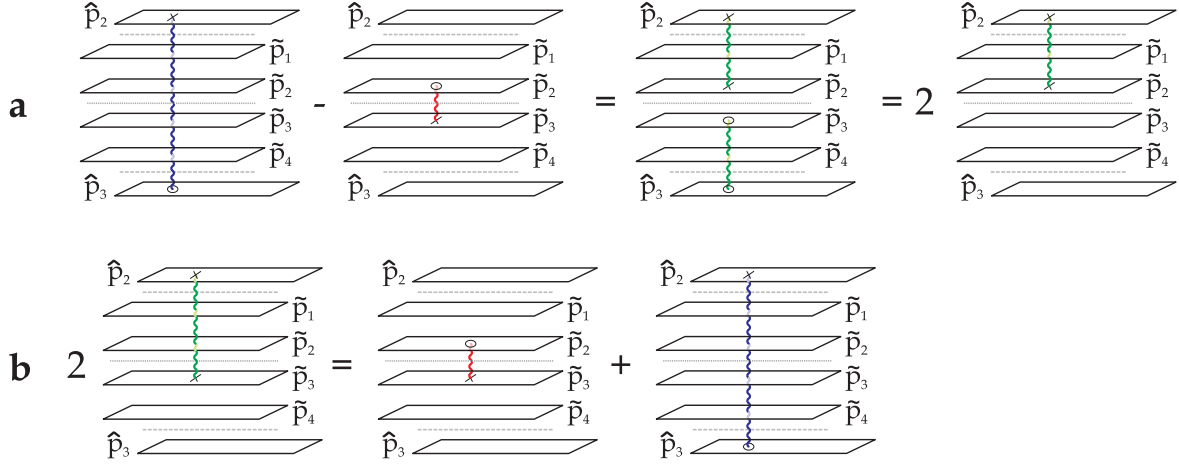}  \caption{\small
  Depiction of equation (\ref{LinearCombiExample}). On top: we see that for symmetric configurations we can obtain the off-sheel fluctuation frequency $\Omega^{\h2\t2}=\Omega^{\t3,\h3}$ from the knowledge of the two $S^5$ and $AdS_5$ frequencies. On bottom: With  this unphysical fluctuation at hand we can compute the fermionic fluctuation frequency $\Omega^{\h2\t3}=\Omega^{\t2\t3}+\Omega^{\h2\t2}$ in terms of the two bosonic fluctuations.   \label{FigureLinearCombi}
 }
  \end{center}
\end{figure}

Consider e.g. the fermionic frequency $\Omega^{\h2 \t3}(y)$. This energy can be thought of as a linear combination of the physical fluctuation $\Omega^{\t 2 \t 3}(y)$ and an unphysical fluctuation $\Omega^{\h2 \t2}(y)$, which in particular does not appear in the table (\ref{Polarisations}) of physical, momentum-carrying polarisations
\beq\label{FermLinearCombi}
\Omega^{\h2 \t3}(y) =  \Omega^{\t 2 \t 3}(y)+ \Omega^{\h2 \t2}(y) \,.
\eeq
Since we are considering symmetric configurations, this unphysical fluctuation energy is identical to $\Omega^{\t3\h3}(y)$, i.e.
\beq
 \Omega^{\h2 \t2}(y) = \Omega^{\t3\h3}(y) \,.
\eeq
As in (\ref{FermLinearCombi}), these unphysical fluctuations can be linearly combined in terms of physical fluctuations
\beq
\Omega^{\h2 \h3}(y)  =\Omega^{\h2 \t2}(y) + \Omega^{\t2 \t3}(y)+ \Omega^{\t3 \h3}(y) \,.
\eeq
Combining all these relations we obtain
\beq\label{LinearCombiExample}
\Omega^{\h2 \t3}(y) = {1\over 2} \left(  \Omega^{\t 2 \t 3}(y) + \Omega^{\h2 \h3}(y)   \right) \,,
\eeq
as depicted in figure \ref{FigureLinearCombi}.

Proceeding in a similar fashion we can derive all frequencies as linear combinations of $\Omega^{\t 2 \t 3}(y)$ and
$\Omega^{\h2 \h3}(y)$. Table (\ref{FreqTable}) summarizes all these relations.

%%%%%%%%%%%%%%%%%%%%%%%%%%%%%%%%%%%%%%%%%
%%%%%%%%%%%%%%%%%%%%%%%%%%%%%%%%%%%%%%%%%

\subsubsection{Final result}

The physical frequencies are labeled by the eight bosonic and eight fermionic polarizations (\ref{Polarisations}), so we can label them by
\beq
\Omega^{ij} \,,\qquad \hbox{where}\quad   i= (\h1, \h2, \t1, \t 2) \qquad   j = (\h3 ,\h4,  \t3 ,\t4) \,.
\eeq
To construct the complete set of off-shell frequencies for a symmetric solution (\ref{SymSol})
in terms of the two fundamental $S^3$ and $AdS_3$ ones $\Omega^{\t2 \t3}(y)$ and $\Omega^{\h2 \h3}(y)$ and their images under $y\rightarrow 1/y$, we first construct by inversion
\beq
\begin{aligned}\label{FreqTable}
\Omega^{\t1\t4} (y) &= -  \Omega^{\t2 \t3}(1/y) + \Omega^{\t2 \t3}(0) \cr
\Omega^{\h1\h4} (y) & = - \Omega^{\h2 \h3}(1/y) -2  \,.
\end{aligned}
\eeq
The remaining frequencies are then obtained by linear combination of these four fluctuation frequencies. In this way we obtain the following concise form for all off-shell frequencies
\beq\label{OmegaFinal}
\Omega^{ij} (y) = {1\over 2} \left(\Omega^{ii'} (y) + \Omega^{j'j} (y) \right) \,,
\eeq
where
\beq
 (\h1, \h2, \t1, \t 2,\h3 ,\h4,  \t3 ,\t4)' = (\h4, \h3, \t 4 , \t 3 ,\h 2, \h 1 , \t 2, \t 1) \,.
\eeq
This generalizes (\ref{LinearCombiExample}), and we have made explicit these linear combinations in appendix B, (\ref{FreqTable}).

%\begin{figure}
%\begin{center}
%  \includegraphics*[width =14cm]{fermion.eps}  \caption{\small
%  Depiction of equation (\ref{LinearCombiExample}). Physical excitations are coloured in red/blue/green. The unphysical intermediate frequency is depicted in grey. \label{fermion}
% }
%  \end{center}
%\end{figure}

In the complete one-loop energy shift (\ref{OneLoopE}) the constant terms in (\ref{FreqTable}) will drop out and thus do not need to be computed. This is particularly obvious, when performing the graded sum  over $\Omega^{ij} (x_n^{ij})$ with the explicit frequencies in (\ref{FreqTable}).

For the general case of not symmetric solutions, we can repeat the above analysis, however the minimal set of required off-shell fluctuation frequencies will generically be larger than two.  It would be interesting to analyse this further.

In the rest of this paper we will consider only $\mathfrak{su}(2)$ solutions which means that only $\t p_2$ (and $\t p_3$) will be connected by square root cuts (outside the unit circle).  For these solutions it is clear that 
\beq
\t p_2=-\t p_3 \,,\qquad \t p_1=-\t p_4\qquad  \text{and} \qquad  \h p_1=\h p_2 =-\h p_3=-\h p_4 \,,    
\eeq
so that we will generically have 6 different frequencies, namely:

\begin{enumerate}
\item{One internal fluctuation corresponding to a pole shared by $\t p_2$ and $\t p_3$ which we denote by
\beq
\Omega_S(y)= \Omega^{\t2  \t3}(y)
\eeq}
\item{Another $S^3$ fluctuation connecting $\t p_1$ and $\t p_4$ 
\beq
\Omega_{\bar S}(y)= \Omega^{\t1  \t4}(y)
\eeq}
\item{Two fluctuations which live in $S^5$ but are orthogonal to the ones in $S^3$,
\beq
\Omega_{S_{\perp}}(y)= \Omega^{\t1  \t3}(y)= \Omega^{\t1  \t4}(y)
\eeq }
\item{Four $AdS_5$ fluctuations
\beq
\Omega_{A}(y)= \Omega^{\h1  \h3 }(y)= \Omega^{\h1  \h4 }(y)=\Omega^{\h2  \h3 }(y)=\Omega^{\h2  \h4 }(y)
\eeq }
\item{Four fermionic excitations which end on either $p_{\t 2}$ or $p_{\t 3}$ (which are the sheets where there are cuts outside the unit circle)
\beq
\Omega_{F}(y)= \Omega^{\h1  \t3 }(y)= \Omega^{\h2  \t3 }(y)=\Omega^{\t2  \h3 }(y)=\Omega^{\t2  \h4 }(y)
\eeq }
\item{Four fermionic poles which end on either $p_{\t 1}$ or $p_{\t 4}$ (which are the sheets where there are cuts inside the unit circle)
\beq
\Omega_{\bar F}(y)= \Omega^{\h1  \t4 }(y)= \Omega^{\h2  \t4 }(y)=\Omega^{\t1  \h3 }(y)=\Omega^{\t1  \h4 }(y) \,.
\eeq }
\end{enumerate}
These fluctuations are depicted in figure \ref{CurveSheets} from left to right.

%%%%%%%%%%%%%%%%%%%%%%%%%%%%%%%%%%%%%%%%%
%%%%%%%%%%%%%%%%%%%%%%%%%%%%%%%%%%%%%%%%%

\section{General $\mathfrak{su}(2)$ two-cut solution}
\label{sec:GeneralTwoCut}

In this section we explain how to compute the fluctuation energies around a general $2$-cut $\mathfrak{su}(2)$ solution\footnote{General two-cut solutions for the $\mathfrak{su}(2)$ Heisenberg magnet were discussed in \cite{Bargheer:2008kj}.} with branch points $a,\bar{a},b,\bar{b}$. We will find out that the fluctuation energies can be obtained
from the surprisingly simple expressions
\beq\label{OmegaTwo}
\begin{aligned}
\Omega_A(y) &= {2 \over y^2-1} \left( 1+ y {f(1) - f(-1) \over f(1) + f(-1)}  \right)
  \cr
\Omega_S(y) &= {4\over f(1) + f(-1)  } \left(  {f(y) \over y^2 -1 } -1 \right)
\,,
\end{aligned}
\eeq
with the remaining fluctuation energies obtained through table (\ref{FreqTable}). Here
\footnote{The proper definition of $f(y)$ is
\beq
f(y)=(2x-a-\bar a)(2x-b-\bar b)\sqrt{\frac{(x-a)(x-\bar a)}{(2x-a-\bar a)^2}}\sqrt{\frac{(x-b)(x-\bar b)}{(2x-b-\bar b)^2}}
\eeq.
 }
\beqa
f(y)\equiv \sqrt{(y-a)(y-\bar a)(y-b)(y-\bar b)} \,.
\eeqa
Note that this is a very simple elegant expression for the off-shell fluctuation energies. All the intricate structure that appears for the on-shell frequencies is hidden in the equation for the pole positions $x_n^{ij}$   (\ref{PolePosition}). 

Let us first review the construction of the quasi-momenta for a two-cut $\mathfrak{su}(2)$ solution. The AdS-quasi-momenta have no cuts, and therefore are rational functions with at most simple poles at $x= \pm1$ and large $x$ asymptotics given by
\beq
p_{\h 1, \h 2} = - p_{\h 3, \h 4} = {2 \pi \mathcal{E} \over x} + O\left({1\over x^2}\right) \,.
\eeq
This determines the AdS quasi-momenta uniquely to be
\beq
p_{\h 1, \h 2} = - p_{\h 3, \h 4} = {2 \pi \mathcal{E} x \over x^2 -1} \,.
\eeq
The derivatives of the sphere quasi-momenta are
\beq\label{Pprime}
p'_{\t2} = - p'_{\t3} = - { \pi \over f(x)}
\left(
{\mathcal{E} f(1) \over (x-1)^2}
+ {\mathcal{E} f'(1) \over x-1} +
{\mathcal{E} f(-1) \over (x+1)^2} +
{\mathcal{E} f'(-1) \over x+1} +2 (\mathcal{J}_1 - \mathcal{J}_2 )
 \right) \,.
\eeq
The remaining sphere quasi-momenta $\t p_1=-\t p_4$ are obtained by the inversion $x\rightarrow 1/x$ as in (\ref{Auto}). 
The first four terms inside the parethesis ensure that the poles of the
quasi-momenta at $x = \pm 1$ are synchronized with the corresponding poles of the AdS quasi-momenta (\ref{VirConst}). Note that $p'_{\tilde{i}}$ is required to have a double pole at $x=\pm1$, with vanishing residue.  The function $1/f(x)$ is needed for the correct inverse square root behaviour close to the branch-points.
The constant terms in the parenthesis are engineered to ensure the correct large $x$ asymptotics (\ref{AsympInfty}).

The moduli of the algebraic curve fix the A and B cycle integrals, and thereby the branch-points. More precisely the moduli are hyperelliptic functions of the branch-points.
Finally to get the quasi-momenta we would have to integrate the meromorphic differential $p' dx$. These last steps will again yield the quasi-momenta as hyperelliptic functions of $x$ and of the branch-points.

In certain instances there can be considerable simplifications due to a degenerate choice of moduli for the curve. This is for example the case for the well-studied symmetric two-cut $\mathfrak{sl}(2)$ solution. Also in the case of the giant magnon, where
 the two cuts are very close, $a \sim b$ and $\bar{a} \sim \bar{b}$, \cite{Minahan:2006bd,Vicedo:2007rp}, we will see that this leads to considerable computational efficiency.

In terms of these unfixed branch-points, however the expression for the derivative of the quasimomenta (\ref{Pprime}) is quite simple as are the expressions for the fluctuation energies anticipated above (\ref{OmegaTwo}).

To discuss the fluctuation frequencies we now perturb the quasi-momenta and fix $\delta p$ by the required asymptotics (\ref{DeltaPAsymp}). We consider only the $(\h2, \h3)$ and $(\t2, \t3)$ fluctuations with $N_{\h2, \h3} = N_{\t2 \t3} =1$, located at $x=z$ and $x=y$ respectively.
The shift in quasi-momenta are
\beq
\begin{aligned}\label{PShift}
\delta p_{\h2} (x; y, z)
& = {\alpha (z) \over x-z}  +
 {\delta \alpha_- \over x-1} + { \delta \alpha_+  \over x+1}    \cr
\delta p_{\t2} (x; y, z)
&= {1 \over f(x)} \left(
-    { f(y) \,\alpha (y) \over x -y} +
   {\delta \alpha_- f(1) \over x-1} + { \delta \alpha_+ f(-1) \over x+1}  - {4 \pi \over \sqrt{\lambda }} \,x  + A
   \right)\,,
\end{aligned}
\eeq
where the asymptotics at large $x$ for $\delta p_{\h2}$, $\delta p_{\t2}$, and also $\delta p_{\h1}$, $\delta p_{\t1}$ obtained by inversion symmetry (\ref{Auto})  fix the constants $\delta \alpha_{\pm}$, $A$ and $\delta \Delta$. We provide the details in Appendix D. The result is
\beq
\delta \Delta =
\Omega_S(y) + \Omega_A(z)  \,,
\eeq
with the notation of (\ref{OmegaTwo}). The remaining constants are summarized in appendix D.

Now that we have found the two off-shell frequencies $\Omega_S$ and $\Omega_A$ we can apply our method from section \ref{sec:QTools} and construct the remaining frequencies as in table (\ref{FreqTable}). In this way we obtain the complete set of fluctuation energies around a generic two cut solution. As an application we will consider in the next section the Giant Magnon solution which corresponds to a particular (singular) limit of the general treatment we considered so far.

Notice also that our simple treatment can be used trivially generalized for $K\ge 3$ cuts.

%%%%%%%%%%%%%%%%%%%%%%%%%%%%%%%%%%%%%
%%%%%%%%%%%%%%%%%%%%%%%%%%%%%%%%%%%%%
%%%%%%%%%%%%%%%%%%%%%%%%%%%%%%%%%%%%%

\section{GM as a two-cut solution}\label{sec:GMTwoCut}

The Giant Magnon solution is a degenerate case of the $2$-cut solution studied in the previous section where the branch points of the algebraic curve are pairwise close.
We will use the explicit formulas (\ref{OmegaTwo}) to compute the frequencies for the giant magnon solution.

In the next subsection we will summarize all the results and then provide the derivations in the subsequent parts.

%%%%%%%%%%%%%%%%%%%%%%%%%%%%%%%%%%%%%%

\subsection{Results}

From the analysis in the last section we have learned that in order to compute the one-loop energy shift (\ref{OneLoopE}), we need the following ingredients:
\begin{itemize}
%\item the classical energy (as we will need to take its derivative as in (\ref{InversionOmega}))
\item the two off-shell $S^3$ and $AdS_3$ fluctuation energies $\Omega_S (y)$ and $\Omega_A(y)$
\item the various quasi-momenta, which are required to determine the position of the physical poles as a function of $n$ (\ref{PolePosition}).
\end{itemize}

Parametrize the branch-points as
\beq\label{BranchPts}
a = X_+ + {\delta \over 2} \,,\qquad
b = X_+ - {\delta  \over 2} \,,
\eeq
and $\bar{a}$ and $\bar{b}$ are complex conjugate to these branch-points, where we denote $X_- = (X_+)^\ast $.
%which we write as
%\beq
%\bar{a} = X_- + {\bar\delta  \over 2} \,,\qquad
%\bar{b} = X_- - {\bar\delta  \over 2} \,.
%\eeq
We will always work up to second order in $\delta$.

Away from the branch-points the two-cuts become indistinguishable, $a \simeq b \simeq X_+$ etc., and the quasi-momenta can be obtained from (\ref{Pprime}) as
\beq
p'_{\t2} (x) = {d \over dx}  \left({ 2 \pi \mathcal{E} x \over x^2-1} +{2 \pi (\mathcal{E } - \mathcal{J} + \mathcal{Q}) \over X_+ - X_- } \log\frac{x-X_+}{x-X_-} \right) \,,
\eeq
where we replaced $\mathcal{J}_1 \rightarrow \mathcal{J}$ and $\mathcal{J}_2 \rightarrow \mathcal{Q}$.
The expression inside the paranthesis is obviously $p_{\t2}(x)$, and the log-cut is the condensate of two cuts with consecutive mode-numbers \cite{Vicedo:2007rp}. The discontinuity by crossing the log-cut is given by $\pi (n+1) - \pi n$ and therefore we can fix the prefactor of the log to be $1/i$, that is to leading order we find
\beq
\mathcal{E } - \mathcal{J} + \mathcal{Q} = {1\over 2 \pi i} (X_+ - X_-)  + O(\delta^2) \,,
\eeq
and therefore
\beq
p'_{\t2} (x) \simeq p_{far}' (x)
\equiv  {d \over dx}  \left({ 2 \pi \mathcal{E} x \over x^2-1} +{1\over i}  \log\frac{x-X_+}{x-X_-} \right) \,,\qquad
|x- X_+|, |x-X_-| \gg \delta 
\,.
\eeq
The quasi-momentum itself is given by 
\beq
p_{far}(x)=\frac{\Delta}{2g}\frac{x}{x^2-1}+\frac{1}{i}\log\frac{x-X_+}{x-X_-}+\tau \la{PnotPrime} \,,
\eeq
where the twist $\tau$ is required to account for the not periodic boundary conditions for the giant magnon and is given by \cite{Gromov:2008ie}
\beq\label{TauDef}
\tau=-p/2=\frac{i}{2}\log\frac{X_+}{X_-} \,.
\eeq
Also, far from the branch-points, $\t p_1(x)=\t p_2(0)+\tau-\t p_2(1/x)$.

Close to the branch-points $a$ and $b$ are given in (\ref{BranchPts}), and the quasi-momentum (\ref{Pprime}) becomes
\beq\label{LogP}
p_{\t2}' (x) \simeq p_{close}' (x) \equiv
{1\over \sqrt{(x-X_+ -{\delta\over 2}) (X_+ -{\delta\over 2} -x )}}   \,,\qquad
|x-X_+| \ll 1\,,
\eeq
where we again used the leading order expression for the energy. Note that up to an overall constant this is obvious, as this is the only function that has the correct branch-cut. Imposing further the same asymptotics for the overlap region $\delta \ll x-  X_+ \ll1 $ as $p_{\t 2}$ in (\ref{LogP}) fixes the overall factor. Alternatively we could fix this constant by imposing $p(b)-p(a)=
\int_a^b p'dx=\pi$ which is precisely what we used above to find the prefactor of the log.

As we will explain below, the classical energy, total filling fraction and momenta of this solution, obtained by integrating the quasi-momenta with suitable measures around the two cuts, will be given by  
\beqa
\Delta-J &=&{g\over i} \left(X_+  - {1\over X_+}  - {\delta^2 \over 8 (X_+)^3} \right) + c.c.  \cr
Q &=&  {g\over i} \left(X_+  + {1\over X_+}  + {\delta^2 \over 8 (X_+)^3} \right)+ c.c.   \label{ChargesExpanded}
\\
P &=& {1\over i } \left(\log X_+  - {\delta^2 \over 16 (X_+)^2}\right) + c.c. \nn \,.
\eeqa
Finally, $\delta$ is fixed by imposing the B-cycle condition $\int_\infty^a  p' = \pi n$, which yields\footnote{The twist $\tau$ is fixed as in the appendix of \cite{Gromov:2008ie}.}
\beq\label{DeltaSquare}
\delta^2 = 16 (X_+ -X_-)^2 \exp\left( - 2 i \tau - i {4 \pi \Delta \over \sqrt{\lambda}}
 {X_+ \over (X_+)^2 -1}\right)\,.
\eeq
These relations allow to parametrize the branch-points $X^{\pm}$ in terms of $\mathcal{Q}$ and $P$, from which then the classical energy $\mathcal{E}$ can easily be computed.

We have determined the off-shell frequencies in the previous section. To obtain the on-shell frequencies $\omega_n$ we compute the positions of
the poles $x_n^{ij}$ from (\ref{PolePosition}) and evaluate them at $x_n$.
There are two case we have to consider. Mainly $x_n$ are situated relatively far from the
branch points of the two cuts and we can expand off-shell frequencies
\beq
\ba
\label{ExpandedOmegas}
\Omega_A(y) &= \Omega^{(0)} (y) -\left(  {y \over y^2 -1 }  {X_+ (X_-^2  - 1)  \over \, 2\,  (X_+^2 -1)(X_+ X_- + 1)^2  }\,  \delta^2 + c.c.\right) \cr
\Omega_S (y) &=
\Omega_A (y)-\(\frac{1}{y-X_+}\frac{X_+-X_-}{4(X_+^2-1)(X_-X_++1)}\delta^2+c.c.\)\,.
\ea
\eeq
The first term is the leading order frequency, as determined in \cite{Gromov:2008ie}, which is
\beq
\Omega^{(0)}(y) = {2\over y^2-1} \left(1- y \, {X_+ + X_- \over X_+ X_- +1} \right)\,.
\eeq
The remaining frequencies are of course determined as in (\ref{OmegaFinal}).

However there are fluctuations corresponding to the
variations of the filling fractions of the two cuts.
These are situated right at the branch points. To compute their
contributions to the 1-loop energy shift we have to expand
$\delta E^{\rm BP}\equiv \frac{1}{2}\Omega^{\t 2\t3}(a)+\frac{1}{2}\Omega^{\t 2\t3}(b)$. 
That leads to
\beq \label{OmBP}
\delta E^{\rm BP}\simeq \Omega^{(0)}(X_+)+\(\frac{1-X_-X_+}{4(X_-X_++1)^2(X_+^2-1)}\delta^2+c.c.\) \,.
\eeq
We will assume that the fluctuations are
situated along the real axis, except the fluctuations
at the branch points, which we will treat separately.

%
%Expanding the frequencies in (\ref{OmegaTwo}) yields
%\beq
%\ba
%\label{ExpandedOmegas}
%\Omega_A (y) &= \Omega^{(0)} (y) -\left(  {y \over y^2 -1 }  {X_+ (X_-^2  - 1)  \over \, 2\,  (X_+^2 -1)(X_+ X_- + 1)^2  }\,  \delta^2 + c.c.\right) \cr
%\Omega_S(y) &=
%\Omega_A(y)-\(\frac{1}{y-X_+}\frac{(X_+-X_-)}{4(X_+^2-1)(X_-X_++1)}\delta^2+c.c.\)
%%\Omega^{(0)} (y) +\left(
%%{\delta^2\over 4 (X_+^2 -1) (X_+ X_- +1)}
%%\left( { X_- - X_+ \over  y- X_+} - {2 y \over y^2 -1} {X_+ (X_-^2 -1)  \over   X_+ X_- +1  }\right) + %c.c\right) \nn
%\ea
%\eeq
%The first term is the leading order frequency, as determined in \cite{Gromov:2008ie}, which is
%\beq
%\Omega^{(0)}(y) = {2\over y^2-1} \left(1- y \, {X_+ + X_- \over X_+ X_- +1} \right)\,.
%\eeq

Now, we have the off-shell fluctuation frequencies, the classical energy as well as the quasi-momenta, and therefore the position of the physical poles, and thus we have all ingredients assembled to compute the one-loop energy shift.

In the next subsection we will derive all the above expressions and subsequently, we will sum up the fluctuation energies to obtain the one-loop energy shift for a generic $Q$-magnon solution.
%For example for a simple giant magnon, with $\mathcal{Q}\ll 1 $, we obtain
%\beq
%\delta E^{1-loop} =  e^{-2 \pi \mathcal{J}  } h(P, \mathcal{J} ) - e^{- {2 \pi \mathcal{J}\over  \sin{p\over 2}}-2} \left(8 \sin^2 {p\over 2} + {16 \sin{p\over 2} \over \pi} \right) \,,
%\eeq
%where the leading exponential term was computed in an exact integral form in \cite{Gromov:2008ie} (see also \cite{Janik:2007wt}).
%In fact 
For the simple giant magnon $\mathcal{Q}\ll 1 $ and the one-loop energy shift organizes as a series in these two exponential \cite{Gromov:2008ie}
\beq
\delta \Delta^{1-loop} = \sum_{n, m} a_{n,m}(P, \mathcal{J}) \,   \Big( e^{- 2 \pi \mathcal{J}}  \Big)^n
\Big(e^{- {2 \pi \mathcal{J}\over \sin{p\over 2}}} \Big)^m \,.
\eeq
In \cite{Gromov:2008ie} we determined the complete set of $a_{n, 0}$ coefficients (see also \cite{Janik:2007wt}), which correct the one-loop shift of the giant magnon in finite volume, by properly summing the leading frequencies as opposed to approximating them by an integral over their momenta. In this paper we determine $a_{1, 1}$, which is the leading correction to the one-loop shift due to the fine-structure of the condensate cut.
Combining the methods in the present paper and  in \cite{Gromov:2008ie}, it should be straight forward to compute $a_{1, n}$.

%%%%%%%%%%%%%%%%%%%%%%%%%%%%%%%%%%%%
%%%%%%%%%%%%%%%%%%%%%%%%%%%%%%%%%%%%

\subsection{Derivations}

\begin{figure}
\begin{center}
  \includegraphics*[width =14cm]{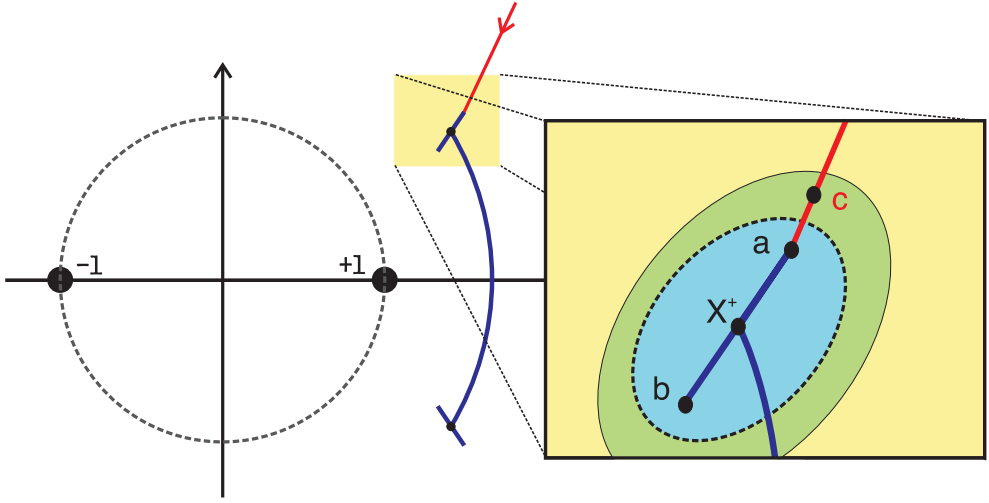}  \caption{\small
Integration regions. \label{picasso}
 }
  \end{center}
\end{figure}

We now provide the details for the results in the last subsection. First let us consider $\delta^2$.
It is determined by fixing the B-cycle integral.  We will compute this integral using different approximations to the quasi-momentum, depending on how far the integration point is from the branch-point
\beq
\pi n=\int_\infty^a p_{\t 2}' = \int_{\infty}^c p'_{far} + \int_{c}^a p'_{close} \,,
\eeq
where $c= X_+ + \epsilon$ is an arbitrary point in the overlapping region $\delta \ll |x-X_+|\ll 1$, i.e. $\delta \ll \epsilon \ll1$. We depicted the integration region in figure \ref{picasso}.
Evaluating the integrals yields
\beq
\pi n \simeq \left[ { 2 \pi \mathcal{E} X_+ \over (X_+)^2-1} +{1\over i}  \log\frac{\epsilon}{X_+-X_-}   + \tau \right]
        + \left[{1\over i}\log {\delta \over 4 \epsilon} \right] \,.
\eeq
Here $\tau$ is the value of the quasi-momenta at infinity. As required  the dependence on $\epsilon$ cancels and we obtain $\delta$ as  function of $X^\pm$ in (\ref{DeltaSquare}).

%%%%%%%%%%%%%%%%%%%%%%%%%%%%%%%%%%%%
%%%%%%%%%%%%%%%%%%%%%%%%%%%%%%%%%%%%

Next we derive the expressions for the charges from the general relations
\beq
\ba
\Delta-J=& { g   \over 2 \pi i}  \oint dx\,  p'(x) \left(x - {1\over x}\right) \cr
Q =& { g   \over 2 \pi i}  \oint dx\,  p'(x)\left(x + {1\over x}\right) \cr
P =& { 1   \over 2 \pi i}  \oint dx\,  p'(x) \log x \,.
\ea
\eeq
For this purpose we write
\beq
\oint dx \, p_{\t2}' (x) f (x) =
\oint dx \, p_{far}'(x) f(x) + \oint dx \left( p_{\t2}'(x) - p_{far}'(x)\right)f(x) \,.
\eeq
The first term obviously yields the leading order charges (\ref{ChargesExpanded}).  The second term can be evaluated by deforming the contour to the region where the integrand is singular, i.e. $x \sim X_+$, where in particular $p_{\t2}$ can be approximated by $p_{close}$
\beq
\oint dx \left( p_{\t2}'(x) - p_{far}'(x)\right)f(x)
\simeq \oint  \left({1\over \sqrt{(x-X_+- {\delta\over 2} )(X_+ - {\delta\over 2}  -x) }} - {i\over x- X_+} \right) f(x)  + c.c. \,,
\eeq
where the contour integral encircles all the poles of the integrand. The integrals can be easily computed and yield (\ref{ChargesExpanded}).

\section{Finite-size correction to the GM} \la{sec5}

The one-loop energy shift is obtained by the weighted sum over all fluctuation frequencies
\beq
\delta\Delta_{1-loop}=\frac{1}{2}\sum_{n,ij}(-1)^{F_{ij}} \Omega_{ij}\(x_n^{ij}\)  \,.
\eeq
To deal with this sum we first split this sum into the fluctuation energies corresponding to a variation of the filling fractions of the two cuts, $\delta E^{\rm BP}$ and the remaining fluctuations. 
To sum the latter we transform the sum over $n$ into an integral with $\cot \pi n$ and then we pass from the $n$ to the $x$ plane using the map (\ref{PolePosition}). Actually as we will explain later there is an additional third contribution coming from fluctuations which got trapped between the two cuts when they collapsed into the log cut. 
This contribution, denoted by $\delta E^{UP}$ is considered in section \ref{sec:UP}. Thus we have
\beq
\delta\Delta_{1-loop}=\frac{1}{2}\sum_{ij}(-1)^F\ccw\oint_{{\cal C}_{\mathbb R}}\Omega^{ij}(y)\cot_{ij}\frac{dy}{2\pi i }+\delta E^{\rm BP}+\delta E^{\rm UP}\;,
\eeq
where
\beq\label{Cotij}
\cot_{ij}\equiv \d_y\log\sin\(\frac{p_i-p_j}{2}\)\;,
\eeq
and the contour $\cal C_{\mathbb R}$ encircles all the fluctuations on the real axis. 

Our goal will be to deform this contour to the unit circle, where the argument of 
the $\cot$ has a large imaginary component
everywhere and the integral can be computed by standard saddle point method.

However, when deforming the contour we will obtain several poles from $\cot_{ij}$ located close to the points $x=X_+,X_-$ and $x=1/X_+,1/X_-$. The contribution from these poles is computed in the next section and is denoted by $\delta E^{\rm PL}$. We find therefore 
\beq
\delta\Delta_{1-loop}=\delta E^{\rm INT}+\delta E^{\rm PL}+\delta E^{\rm BP}+\delta E^{\rm UP}\;,
\eeq
where
\beq
\delta E^{\rm INT}=\cw\oint_{{\cal C}_{\mathbb U}}\(\frac{1}{2}\sum_{ij}(-1)^F\Omega^{ij}(x)\cot_{ij}\)\frac{dx}{2\pi i }  \,. \la{unitc}
\eeq
Notice that since we already dealt with the zero mode contribution $\omega_{BP}$ separately we can (and will) use the far away quasi-momenta (\ref{PnotPrime}) in the rest of the paper. In the following four sections we will consider each of these four contributions in detail.

The splitting of the one-loop shift into a unit circle contribution plus the rest was proposed in \cite{Gromov:2007cd} where the unit circle contribution was analyzed and related with the Hernandez-Lopez phase \cite{Hernandez:2006tk}. In \cite{Gromov:2007ky} the remaining contribution was considered around general non-singular classical curves and matched with the usual finite size corrections, known as anomalies, appearing in the Beisert-Staudacher equations \cite{Beisert:2005fw}.

%%%%%%%%%%%%%%%%%%%%%%%%%%%%%%%%%%

\subsection{Extracting poles}

We now determine the positions of the poles mentioned above.
Consider first the polarization $(\t2,\t3)$. We have 
\beq
\exp(-i\t p_2+i\t p_3)=\exp\(-i\frac{x\Delta}{g(x^2-1)}-2i\tau\)\frac{(x-X_-)^2}{(x-X_+)^2}\,,
\eeq
so we will have an obvious pole from (\ref{Cotij}) at $x=X^+$ but we will also have some less trivial poles 
if the denominator in (\ref{Cotij}) vanishes, i.e. for $\exp(-i\t p_2+i\t p_3)=1$,
\beq
\exp\(-i\frac{x\Delta}{g(x^2-1)}-2i\tau\)\frac{(x-X_-)^2}{(x-X_+)^2}=1\,.
\eeq
The first factor is exponentially small. When $x\sim X_+$
the exponent is of order $\delta^2$ as one can see from (\ref{DeltaSquare}).
However we can compensate that if the second factor diverges.
To be able to compensate the exponential supression we will
require that $x-X_+\sim \delta$. What one finds is the poles at
$x-X_+=\epsilon^\pm_1$, where
\beq
\ba
\epsilon^\pm_1 
&= \pm\frac{\delta}{4}+ \frac{\delta^2}{16} \(\frac{1}{X_+-X_-}+i\frac{\Delta}{2g}\frac{X_+^2+1}{(X_+^2-1)^2}\)\cr
&\pm\frac{\delta^3}{64}
\(\frac{1}{(X_--X_+)^2}-\frac{3\Delta^2(X_+^2+1)^2}{8g^2(X_+^2-1)^4}
+\frac{i\Delta}{2g}\frac{2 X_+^4+X_- X_+^3-3X_+^2+3X_-X_+-3}{(X_+-X_-)(X_+^2-1)^3}\)
+\OO\(\delta^4\)
\ea
\eeq
Proceeding in the same way for the different polarizations we would find the position of all existing poles.  We have summarized all poles, and whether they are physical or unphysical (around $X_+$ or $1/X_+$, respectively) in table \ref{tab:1}. In Appendix E we listed the explicit values of the small deviations $\epsilon_j$. 

%%%%%%%%%%%%%

\begin{table}[h]
\begin{center}
\begin{tabular}{@{}cll@{}}
  \toprule
  \bf Polarization & \bf Poles around $X_+$ & \bf Poles around $1/X_+$ \\ \midrule
  $A\times 4$ &  &  \\ \midrule
  $F\times 4$ & $x-X_+=0,\epsilon_{3}$ &  \\ \midrule
  $\bar F\times 4$ &  & $1/x-X_+=0,\epsilon_{3}$ \\ \midrule
  $S$ & $x-X_+=\epsilon^-_{1},0,\epsilon^+_{1}$ &  \\ \midrule
  $\bar S$ &  & $1/x-X_+=\epsilon^-_{1},0,\epsilon^+_{1}$ \\ \midrule
  $S_\perp\times 2$ & $x-X_+=0,\epsilon_{2}$ & $1/x-X_+=0,\epsilon_{2}$ \\ \bottomrule
\end{tabular}
\end{center}
\caption{Poles of different $\cot_{ij}$ in the upper half plane close to the logarithm branch points }\la{tab:1}
\end{table}

%%%%%%%%%%%%%%

In summary, the contribution to the contour integral from these singularities is
\beq\label{Epl}
\delta E^{\rm PL}=\(
\frac{e^{i\tau}}{(X_-X_++1)(X_+^2-1)}+
\frac{2-X_+(X_-+X_+)}{(X_-X_++1)(X_+^2-1)^2}
+\frac{i\Delta}{4g}\frac{(X_--X_+)(X_+^2+1)}{(X_-X_++1)(X_+^2-1)^3}
\)\frac{\delta^2}{4}+c.c. \,.
\eeq
which for small $Q$ values becomes
\beq\label{PlQ}
\delta E^{\rm PL}\simeq
8 e^{-\frac{J}{2g\sin\frac{p}{2}}-2} \sin^2\frac{p}{2} \,.
\eeq

\subsection{Unphysical fluctuations}
\label{sec:UP}

\begin{figure}[ht]
\begin{center}
  \includegraphics*[width =12cm]{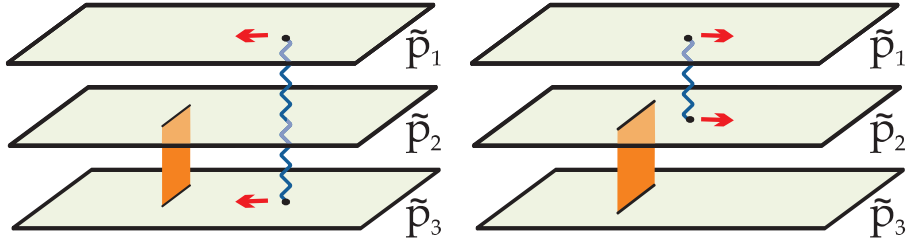}  
  \caption{When increasing the filling fraction of a cut, the fluctuation could pass through the cut and reappear uniting different sheets. The physical fluctuation $\t2\t3$ could become the unphysical one $\t1\t2$.\label{UnFl}
 }
  \end{center}
\end{figure}

Consider a general finite gap solution.
Let us assume first that all filling fractions
are sufficiently small. By other words we are dealing with a slightly deformed BMN
curve. Then we know that the equation
\beq
p_i(x_n^{ij})-p_j(x_n^{ij})=2\pi n\la{xijn}
\eeq
for a physical pair $(ij)$
always has a solution\footnote{In fact one should add twists to ensure this statement.}.
When we gradually start increasing the filling fractions, the cuts become bigger and at some point
a cut could collide with some $x_n$.
After this point we will not be able to find solutions to (\ref{xijn}) for some values of $n$.
This however does not imply any non-analyticity of the fluctuation
energies $\Omega^{ij}(x^{ij}_n)$
as a function of the filling fractions and we can analytically continue the fluctuation energies
below this point. What happens is that the fluctuation $x_n$ passes through a cut and afterwards is connecting two different sheets. This will generically yield unphysical fluctuations. We have depited this process in figure \ref{UnFl}.

Indeed for each missing solution of $\eq{xijn}$ one could find the corresponding unphysical fluctuation.
We conclude that we also have to consider all possible solutions of $\eq{xijn}$ for unphysical pairs $(ij)$.

In the calculation above we have taken into account only physical fluctuations. However there are
$2+4$ unphysical fluctuations $(\t1,\t2),(\t3,\t4)$ and $(\h 1\t2),(\h2\t2),(\h3\t3),(\h4\t3)$, which by the above reasoning we also need to take into account. We denote these fluctuations by $S_u$ and $F_u$
\beqa
\Omega_{S_u}(x)&=&\frac{\Omega_{\bar S}(x)-\Omega_S(x)}{2}+c\\
\Omega_{F_u}(x)&=&\frac{\Omega_A(x)-\Omega_S(x)}{2}+c\,,
\eeqa
where the specific values of the constant $c$ and of the position of these fluctuations $x^{S_u}$ and $x^{F_u}$ are collected in Appendix E.

%%%%%%%%%%%%%%%%%%%%

%%%%%%%%%%%%%%%%%%%%%%%

Combining this together with the branch-point contribution (\ref{OmBP}) one obtains
\beq\label{Eun}
\delta E^{\rm UP}+\delta E^{\rm BP}=\delta E^{BP}+\frac{2\Omega_{S_u}(0)-4\Omega_{F_u}(x^{F_u})}{2}=
\frac{1-\sqrt{X_-/X_+}}{4(X_-X_++1)(X_+^2-1)}\delta^2+c.c. \,,
\eeq
where in particular the leading order term correctly cancels!
In the $\mathcal{Q}\to 0$ limit we obtain for this combined contribution
\beq
\delta E^{\rm UP}+\delta E^{\rm BP}\simeq-8 e^{-\frac{J}{2g\sin\frac{p}{2}}-2}\sin^2\frac{p}{2} \,.
\eeq
This contribution precisely cancels the contribution of  $\delta E^{\text PL}$ in (\ref{PlQ}). Thus, for the simple giant-magnon solution the only contribution is given by the integral over the unit circle (\ref{unitc})

%%%%%%%%%%%%%%%%%%%%%

\subsection{On the unit circle}
In the previous two section we took into account the extra poles in the complex $x$ plane, the branch-point fluctuations and the unphysical excitations. For a general Dyonic magnon these contributions are given by (\ref{Epl}) added to (\ref{Eun}) while for a simple giant-magnon this sum vanishes.

In this section we consider the remaining contribution given by the integral (\ref{unitc}) over the unit circle. There are three contributions into which this integral is naturally split. 
On the upper/lower half of the unit circle we have 
\beq
\cot\(\frac{p_i-p_j}{2}\)=\pm i\(1+2 e^{\mp i (p_i-p_j)}+\dots \) \la{cote} \,,
\eeq
while the fluctuation energies are given by
\beq
\Omega_{ij}(y)=\Omega^{(0)}(y)+\delta \Omega_{ij}(y) \,.\la{flucte}
\eeq
Thus the we can pick the leading term in (\ref{cote}) times the leading term in (\ref{flucte}) to get 
\beq
\delta E^{\rm INT,(0)}=\cw\oint_{{\cal C}_{\mathbb U}^+} \frac{dy}{2 i}(-1)^{F_{ij}} \partial_y\Omega^{(0)}(y) \frac{p'_i-p'_j}{2\pi} \,,
\eeq
where the integral goes over the upper half of the unit circle from $x=-1$ to $x=+1$. Since $\sum_{i=1}^4 \t p_i-\h p_i=0$ this contribution vanishes and therefore the one-loop shift around the infinite volume giant magnon is zero.
We are therefore left with the exponentially suppressed contributions.

The  second contribution comes from picking the subleading term in (\ref{flucte}) and the leading value in (\ref{cote}). This gives
\beq
\ba\label{Eint}
\delta E^{{\rm INT},(1)}
%&=\cw\oint_{{\cal C}_{\mathbb U}}\(\frac{1}{2}\sum_{ij}(-1)^F\Omega^{ij}(x)\cot_{ij}\)\frac{dx}{2\pi i } \cr
&\simeq 2\cw\oint_{{\cal C}_{\mathbb U}^+}\frac{h(x)-h(1/x)/x^2+g(x)}{(X_+^2-1)(X_+X_-+1)}\frac{dx}{2\pi i}+c.c.\cr
&=\frac{i\delta^2}{4\pi}\[\frac{X_+-X_-}{(X_+X_-+1)(X_+^2-1)^2}+\frac{(X_-^2-1)({\rm arccoth}\,X_+-{\rm arccoth}\,X_-)}{(X_+^2-1)(X_+^2X_-^2-1)}\]+c.c.\,,
\ea
\eeq
where
\beq
\ba
h(x)&=\frac{\delta^2}{16}\[\frac{X_--X_+}{(x-X_+)^2}+\frac{X_--2X_++X_-X_+^2}{X_+(X_-X_+-1)}\(\frac{1}{x-X_+}-\frac{1}{x-X_-}\)\]\cr
g(x)&=\frac{\delta^2}{8}\frac{(X_+-X_-)^2}{(x X_+-1)(x X_--1)X_+} \,.
\ea
\eeq
Expanding this result in the $\mathcal{Q}\to 0$ limit we obtain
\beq\label{Eintone}
\delta E^{{\rm INT}, (1)}\simeq
16 e^{-\frac{J}{2g\sin\frac{p}{2}}-2}
\(
\frac{g\sin^3\frac{p}{2}}{Q}-
\frac{\sin\frac{p}{2}}{\pi}
\) \,.
\eeq
Notice that this contribution is singular in the $\mathcal{Q}\rightarrow 0$ limit. This singularity will cancel however with the third contribution we will now analyze.

Finally we have the contribution coming from picking the leading term in (\ref{flucte}) multiplied by the subleading term in (\ref{cote}). This was the contribution analyzed in \cite{Gromov:2008ie} and \cite{Janik:2007wt}. This gives
\beqa
&&\delta E^{{\rm INT}, (2)}=\cw\oint_{U^+} \frac{dx}{2\pi i} \partial_x \Omega_0 \(e^{-i\tau} \frac{x-X_-}{x-X_+}+e^{-i\tau} \frac{x-1/X_+}{x-1/X_-}-2\)^2 e^{-\frac{ix\Delta}{g(x^2-1)}} \,,
\eeqa
which in the small $\mathcal{Q}$ limit is divergent and becomes
\beq\label{Eexp}
\ba
\delta E^{{\rm INT}, (2)}
\simeq \,&V.P. \ \cw \oint_{U^+} \frac{dx}{2\pi i} \partial_x \Omega_0 \(
2\frac{xX_+-1}{x-X_+}-2
\)^2 e^{-ix\frac{J+4g\sin\frac{p}{2}}{g(x^2-1)}}\\
 &+e^{-\frac{J}{2g\sin\frac p2}-2}\(-\frac{16g\sin^3\frac p2}{Q}+\frac{4iJ\cos\frac{p}{2}}{g}-8i\sin\frac{p}{2}+8i\sin p\)\,,
\ea
\eeq
where $V.P.$ stands for the principal value of the integral. 

Finally, we can combine (\ref{Eintone}) and (\ref{Eexp}) to obtain the final result
\beq\label{GMFinal}
\ba
\delta \Delta_{1-loop}&\simeq 
V.P.\cw \oint_{U^+} \frac{dx}{2\pi i} \partial_x \Omega_0 \( 2\frac{xX_+-1}{x-X_+}-2
\)^2 e^{-ix\frac{J+4g\sin\frac{p}{2}}{g(x^2-1)}}\cr
&\qquad +e^{-\frac{J}{2g\sin\frac p2}-2}\(-\frac{16\sin\frac p2}{\pi}+\frac{4iJ\cos\frac{p}{2}}{g}-8i\sin\frac{p}{2}+8i\sin p\) \,.
\ea
\eeq
We will show in the next section that this is in precise agreement with the 
 $F$ and $\mu$ terms of the L\"uscher-Klassen-Melzer formulas!
Note that the expression above is real by construction and the divergence at $\mathcal{Q}=0$ has cancelled among the various contributions.

%%%%%%%%%%%%%%%%%%%%%%%%%%%%%%%%%%%%%%%%%%%

\subsection{Combined energy shift for a generic Dyonic magnon}

Notice that we are by no means obliged to take the simple magnon magnon and our previous formulas are absolutely general and also yield the finite size $1$-loop shift around a generic Dyonic magnon. Combining all the contributions computed in the previous sections we get
\beqa\label{Combi}
\delta \Delta_{1-loop}&=&
\cw\oint_{U^+} \frac{dx}{2\pi i} \partial_x \Omega_0 \(e^{-i\tau} \frac{x-X_-}{x-X_+}+e^{-i\tau} \frac{x-1/X_+}{x-1/X_-}-2\)^2 e^{-\frac{ix\Delta}{g(x^2-1)}}  \\
&+&\( \frac{\delta^2}{4(X_-X_++1)(X_+^2-1)^2}\[ 1-X_-X_++i\frac{X_+-X_-}{\pi} \nn
-i\,\frac{\Delta}{4g}\frac{(X_+^2+1)(X_+-X_-)}{X_+^2-1}\right. \right. \\ \nn
&+&\left.\left.i\frac{(X_-^2-1)(X_+^2-1)}{2\pi(X_-X_+-1)}\log\(\frac{(X_++1)(X_--1)}{(X_+-1)(X_-+1)}\)
\]+c.c.\)\,.
\eeqa

\section{L\"uscher-Klassen-Melzer formulas}
\label{sec:Luscher}

Finally we compute the finite-size correction (\ref{GMFinal}) using the L\"uscher-Klassen-Melzer formulas \cite{Luscher:1985dn, Luscher:1986pf, Klassen:1990ub, Ambjorn:2005wa, Janik:2007wt, Gromov:2008ie, Heller:2008at}. 
There are two contributions, the $F$- and the $\mu$-term
\beqa
&&\delta\epsilon^F_a=-V.P.\int_{\mathbb R}\frac{dq}{2\pi}\(1-\frac{\epsilon'(p)}{\epsilon'(q^*(q))}\)
e^{-iq^*(q)L}\sum_b(-1)^{F_b}S_{ba}^{ba}(q^*(q),p)
\\
&&\delta \epsilon^{\mu}_a =
- i \left(1 - {\epsilon'(p) \over \epsilon'(\tilde{q}^*)} \right)
e^{- i\tilde{q}^* L} \hbox{Res}_{q = \tilde{q}}  \left(\sum_b (-1)^{F_b}  S_{ba}^{ba} (q_*(q) , p) \right) \,,
\eeqa
which describe the corrections to the dispersion relation of a single magnon with momentum $p$ due to virtual particles running in the loop, and bound state formation, respectively. 
We have used the notation for the on-shell momentum
\be
q^2 + \epsilon(q_*)^2 = 0  \,,
\ee
and $\tilde{q}$ denotes the Euclidean energy of the bound state. 
Inserting the all-loop AdS/CFT S-matrix  \cite{Beisert:2005tm, Beisert:2006ib, Beisert:2006ez, Arutyunov:2006yd}, one can expand to arbitrary order and obtain the leading-volume correction. 

Through a trivial change of variables, the F-term can be written as \cite{Gromov:2008ie}
\beq\label{Fterm}
\delta \epsilon^F = 
V.P. \cw\oint_{\mathbb{U}^+} \frac{dx}{2\pi i}\,\d_x\Omega_0(x) \, e^{-4\pi\frac{iJ}{\sqrt{\lambda}}\frac{x}{x^2-1}}
 e^{-4\pi \frac{i(\Delta-J)}{\sqrt{\lambda}}\frac{x}{x^2-1}}
 \(2\, \frac{x-X_-}{x-X_+}\sqrt\frac{X_+}{X_-}-2\)^2 \,,
 % \sum_{b} (-1)^{F_b} S_{ba}^{ba}(q^*(q),p)\,,
\eeq
where 
\beq
\Delta = J + {\sqrt{\lambda} \over \pi} \sin{p\over 2} \,.
\eeq
In the limit of $\mathcal{Q}\rightarrow 0$, $X_+ \sim 1/X_-$  and thus the F-term agrees precisely with the first line in (\ref{GMFinal})!

For the $\mu$-term we have to evaluate the residue at the bound states as done in  \cite{Janik:2007wt}, to subleading order. Since the computation is exactly as done in this paper we omit the details. There are
three contributions, which arise from poles up to the one-loop dressing factor, the effect from the one-loop dressing factor and the higher-loop contributions, respectively. In summary we obtain
\beq
\delta \epsilon^\mu=e^{-\frac{2\pi J}{\sqrt{\lambda}\sin\frac{p}{2}}}  \delta_1 \delta_2 \delta_3 \,,
\eeq
where
\beqa
\delta_{1}&=&-4 g \sin^3\frac{p}{2}+i \({J \over g} \cos\frac{p}{2}-2\sin\frac{p}{2}+\sin p\) +\mathcal{O}\(\frac 1g\) \\
\delta_{2}&=&\frac{1}{2}+\frac{1}{g}\(\frac{1}{2\pi\sin^2 \frac
p2}-\frac{i\cos\frac p2}{4\sin^2 \frac p2}\)+\mathcal{O}\(\frac 1{g^2}\)\\
\delta_{3}&=&\frac{8}{e^2} +\mathcal{O}\(\frac 1{g^2}\) \,,
\eeqa
so that the $\mu$-term up to this order is
\beq
\delta\epsilon_\mu=
-e^{-\frac{2 \pi  J}{\sqrt{\lambda} \sin\frac p2}-2}\(16 g \sin^3\hp+\frac{16}{\pi}\sin\hp -4i\(\J\cos\hp-2\sin\hp+2\sin p\)\) 
 + \mathcal{O}\left({1\over g}\right)\,.
\eeq
The leading $\mathcal{O}(g)$ contribution to the $\mu$-term is precisely the one in \cite{Janik:2007wt}. The subleading terms are in complete agreement with the corrections appearing in the second line of our result (\ref{GMFinal}). 
Thus we have successfully demonstrated the agreement of our result (\ref{GMFinal}) with the L\"uscher-Klassen-Melzer approach of computing finite-size effects. As we emphasized already in the introduction, we strongly believe that this efficient quantization method that we developed in this paper will be useful in systematic studies of the  finite-size effects for strings in $AdS_5 \times S^5$. In particular it would be interesting to reproduce from a L\"uscher like approach the full result (\ref{Combi}) for the finite size corrections to the Dyonic magnon.

%Mention F-term is tricky. Computation of mu term. Computation of next order of F-term could be funny.

%%%%%%%%%%%%%%%%%%%%%%%%%%%%%%%%%%%%%
%%%%%%%%%%%%%%%%%%%%%%%%%%%%%%%%%%%%%
%%%%%%%%%%%%%%%%%%%%%%%%%%%%%%%%%%%%%

%\section{Discussion}

\subsection*{Acknowledgments}

We thank V.~Kazakov for discussions. NG was partially supported by RSGSS-1124.2003.2, by RFBR grant 08-02-00287 and ANR grant INT-AdS/CFT (contract ANR36ADSCSTZ).
The work of SSN is supported by John A. McCone Postdoctoral Fellowship. SSN would like to thank the IPMU and the University of Tokyo for hospitality during some stages of this work.
PV is funded by the Funda\c{c}\~ao para a Ci\^encia e Tecnologia fellowship {SFRH/BD/17959/2004/0WA9}. PV would like to thank KITP for warm hospitality. PV would like to thank California Institute of Technology where part of this work was done for warm hospitality. 

\newpage

%%%%%%%%%%%%%%%%%%%%%%%%%%%%%%%%%%%%%
%%%%%%%%%%%%%%%%%%%%%%%%%%%%%%%%%%%%%
%%%%%%%%%%%%%%%%%%%%%%%%%%%%%%%%%%%%%

\setcounter{section}{0}
\setcounter{subsection}{0}

%%%%%%%%%%%%%%%%%%%%%%%%%%%%%%%%%%%%%
%%%%%%%%%%%%%%%%%%%%%%%%%%%%%%%%%%%%%
%%%%%%%%%%%%%%%%%%%%%%%%%%%%%%%%%%%%%

\appendix{Classical Curve and Fluctuations}

\subsection{Properties of the Classical Curve}
\label{sec:AppendixClassics}

In this appendix we summarize the properties of the classical algebraic curve.
The monodromy matrix $\Omega (x)$ is uni-modular, STr$(\Omega (x)) =1$, so that
\beq\label{UniMod}
\sum_{i=1}^4 \left( \tilde{p}_{i} - \hat{p}_{i} \right)  \in  2 \pi {\mathbb Z}  \,.
\eeq
The eigenvalues of the monodromy matrix can have branch-cuts along which the quasi-momenta can jump as
\beq\label{AlgCurveEq}
 p_i^+-  p_j^-=2\pi n_{ij} \,, \qquad  x\in \mathcal{C}_{n}^{ij}  \,,
\eeq
for the combination of sheets
\beq
\la{range} i=\t 1,\t 2,\h 1,\h 2 \,, \quad j=\t 3,\t 4,\h 3,\h 4 \,.
\eeq
We use the notation $p_{\t i} = \tilde{p}_i$ etc. interchangeably. Note that for $ij$ of the same type, $\mathcal{C}_{n}^{ij}$ are macroscopic bosonic cuts, whereas if they are of different type these are fermionic poles.

The quasi-momenta can have branch-cuts and poles in the $x$-plane. The latter are located at $x =\pm 1$ with residues that are related in particular due to the Virasoro constraint
\beq\label{VirConst}
\{\h p_1,\h p_2,\h p_3,\h p_4|\t p_1,\t p_2,\t p_3,\t p_4\}
=
\frac{\{  \alpha_\pm,  \alpha_\pm,  \beta_\pm,  \beta_\pm|  \alpha_\pm,  \alpha_\pm,  \beta_\pm,  \beta_\pm\}}{x\pm 1}  + O(1)\,.
\eeq
At $x= \infty$ the BPR connection $A(x)$ reduces simply to the Noether current, wherefore the quasi-momenta
in this limit are related to the global $\mathfrak{psu} (2,2|4)$ charges
\beq \label{AsympInfty}
\(\bea{c}
\h p_1\\
\h p_2\\
\h p_3\\
\h p_4\\   \hline
\t p_1\\
\t p_2\\
\t p_3\\
\t p_4\\
\eea\) = \frac{2\pi }{x}
\(\bea{l}
+\mathcal{E}-\mathcal{S}_1+\mathcal{S}_2 \\
+\mathcal{E}+\mathcal{S}_1 -\mathcal{S}_2 \\
-\mathcal{E}-\mathcal{S}_1  -\mathcal{S}_2 \\
-\mathcal{E}+\mathcal{S}_1 +\mathcal{S}_2 \\ \hline
+\mathcal{J}_1+\mathcal{J}_2-\mathcal{J}_3  \\
+\mathcal{J}_1-\mathcal{J}_2+\mathcal{J}_3 \\
-\mathcal{J}_1+\mathcal{J}_2 +\mathcal{J}_3 \\
-\mathcal{J}_1-\mathcal{J}_2-\mathcal{J}_3
\eea\)   + O\left({1 \over x^2}\right)\,.
\eeq
The algebra $\mathfrak{psu}(2,2|4)$ further has an automorphism which acts also on the monodromy matrix and imposes the following relations for the quasi-momenta
\beqa\label{Auto}
\t p_{1,2}(x)&=&-\t p_{2,1}(1/x)-2\pi m\cr
\t p_{3,4}(x)&=&-\t p_{4,3}(1/x)+2\pi m\cr
\h p_{1,2,3,4}(x)&=&-\h p_{2,1,4,3}(1/x)\,.
\eeqa

Each classical solution is specified by the following data:
bosonic branch-cuts (connecting $\t i$ and $\t j$ or $\h i$ and $\h j$) and fermionic poles (connecting $\t i$ and $\h j$), which are specified by the quasi-momenta $p_i$, satisfying (\ref{UniMod}, \ref{AlgCurveEq}, \ref{VirConst}, \ref{AsympInfty}, \ref{Auto}). Each cut carries a mode number, $n_{ij}$ in (\ref{AlgCurveEq}) and furthermore a filling fraction
\beq
S_{ij}=\pm\,\frac{\sqrt{\lambda}}{8\pi^2i}\oint_{\mathcal{C}_{ij}} \(1-\frac{1}{x^2}\) p_i(x) dx \label{FillingFrac}\,.
\eeq

%%%%%%%%%%%%%%%%%%%%%%%%%%%%%%%%%%%%%
%%%%%%%%%%%%%%%%%%%%%%%%%%%%%%%%%%%%%
%%%%%%%%%%%%%%%%%%%%%%%%%%%%%%%%%%%%%

\subsection{Properties of Fluctuations}
\label{sec:AppendixFluctus}

As for classical quasi-momenta, we obtain constraints on the asymptotics for $\delta p_i$, which should include a sum over all fluctuations $f^{(i,j)}$
\beq
\delta p_i \sim \sum_{(i, j)} f^{(i,j)}
            = \sum_{(i, j)}\epsilon_i N_n^{ij} {\alpha (x_n^{ij}) \over x- x_n^{ij}} \,.
\eeq
The signs appearing here and in (\ref{PolePosition}) are
\beq\label{FluctuSigns}
1= \epsilon_{\h 1} = \epsilon_{\h 2} = - \epsilon_{\h 3} = -\epsilon_{\h 4}
 = -\epsilon_{\t 1} = -\epsilon_{\t 2} = \epsilon_{\t 3} = \epsilon_{\t 4} \,,
\eeq
and we defined
\beq\label{AlphaDef}
\alpha (x) = {4 \pi \over \sqrt{\lambda}} {x^2 \over x^2 -1} \,.
\eeq
The corrections to the quasi-momenta can again have poles at $x =\pm 1$ which have to be correlated via the Virasoro constraint as
\beq\label{DeltapVirConst}
\{\delta \h p_1,\delta\h p_2,\delta\h p_3,\delta\h p_4|\delta\t p_1,\delta\t p_2,\delta\t p_3,\delta\t p_4\}
=
\frac{\{  \delta\alpha_\pm,\delta  \alpha_\pm, \delta \beta_\pm, \delta \beta_\pm|
\delta\alpha_\pm, \delta \alpha_\pm, \delta \beta_\pm,  \delta\beta_\pm\}}{x\pm 1}   + O(1)\,.
\eeq

The asymptotics at $x= \infty$ of the quasi-momenta (\ref{AsympInfty}) results in constraints on
the asymptotics of the deformed quasi-momenta
\beq
   \(\bea{c}
\delta \h p_1\\
\delta\h p_2\\
\delta\h p_3\\
\delta\h p_4\\  \hline
\delta\t p_1\\
\delta\t p_2\\
\delta\t p_3\\
\delta\t p_4\\
\eea\)
=   \frac{4 \pi}{x \sqrt{\lambda}  }\!
\(\bea{rrl}
+\delta \Delta/2&+N_{\h 1\h 4}+N_{\h 1 \h 3}&+N_{\h 1\t 3}+N_{\h1\t 4}\\
+\delta \Delta/2&+N_{\h 2\h 3}+N_{\h 2\h 4}&+N_{\h2\t 4}+N_{\h2\t 3} \\
-\delta \Delta/2&-N_{\h 2\h 3}-N_{\h 1 \h 3}&-N_{\t 1\h3}-N_{\t 2\h3} \\
-\delta \Delta/2&-N_{\h 1\h 4}-N_{\h 2\h 4}&-N_{\t 2\h4}-N_{\t 1\h4} \\ \hline
&- N_{\t1 \t4}- N_{\t 1\t 3} &-N_{\t 1\h3}-N_{\t 1\h4} \\
&- N_{\t 2 \t3}- N_{ \t 2 \t4}                         &-N_{\t 2\h4}-N_{\t 2\h3}\\
&+ N_{ \t2 \t3}+ N_{ \t1 \t3}&+N_{\h1\t 3}+N_{\h2\t 3}\\
&+ N_{ \t1 \t4}+ N_{ \t 2 \t4}             &+N_{\h2\t 4}+N_{\h1\t 4}
\eea\) + O\left({1\over x^2} \right) \,. \label{DeltaPAsymp}
\eeq
where $\delta\Delta$ is given by
\beq
\delta \Delta = \sum_{ij,n} N_{ij}^n \Omega_{ij}^n
\eeq
The important point is that these are related to the energy shift $\delta E$.

So far we covered all the constraints that follow from the asymptotics of the classical
quasi-momenta. In addition, the fluctuations will backreact upon the classical cuts and close to the branch-points (or cut-endpoints) we impose for $p_i \sim \sqrt{(x-a)}$ close to the branch-point $x=a$
\beq
\delta p_i \sim {d \over d x} p_i \,,
\eeq
etc.
Solving these constraints in particular fixes $\delta E$, which is the desired one-loop energy shift.

%%%%%%%%%%%%%%%%%%%%%%%%%%%%%%%%%%%%%
%%%%%%%%%%%%%%%%%%%%%%%%%%%%%%%%%%%%%

\appendix{Complete set of frequencies}

The complete set of frequencies written in terms of the basis frequencies $\Omega^{\t2 \t3} (y)$ and $\Omega^{\h2 \h3}(y)$ are
\beq
\begin{aligned}\label{FreqTable}
\Omega^{\t1\t4} (y) &= -  \Omega^{\t2 \t3}(1/y) +\Omega^{\t2\t3}(0)\cr 
\Omega^{\t2 \t4} (y) =
\Omega^{\t1\t3}(y)  &= {1\over 2} \left( \Omega^{\t2 \t3}(y) + \Omega^{\t1\t4} (y) \right)
                               = {1\over 2} \left( \Omega^{\t2 \t3}(y) - \Omega^{\t2\t3} (1/y)+\Omega^{\t2\t3}(0) \right)
                                 \cr
 \Omega^{\h1\h4} (y) & = - \Omega^{\h2 \h3}(1/y) -2 \cr
\Omega^{\h2 \h4} (y) =
\Omega^{\h1 \h3} (y) &= {1\over 2}  \left( \Omega^{\h2 \h3}(y) + \Omega^{\h1\h4} (y) \right)
				    = {1\over 2}  \left( \Omega^{\h2 \h3}(y) - \Omega^{\h2\h3} (1/y) \right) -1 \cr
\Omega^{\h2\t4}(y) =
\Omega^{\t1 \h3} (y) &= {1\over 2} \left(\Omega^{\h2 \h3}(y)+ \Omega^{\t1\t4} (y)  \right)
				   = {1\over 2} \left( \Omega^{\h2 \h3}(y)- \Omega^{\t2 \t3}(1/y)+\Omega^{\t2\t3}(0) \right)  \cr				
\Omega^{\t2\h4}(y)=
\Omega^{\h1 \t3} (y) &= {1\over 2} \left( \Omega^{\t2\t3} (y) + \Omega^{\h1 \h4}(y)\right)
				 = {1\over 2}  \left( \Omega^{\t2\t3} (y) - \Omega^{\h2 \h3}(1/y)\right)   -1 \cr
\Omega^{\t1 \h4} (y) =
\Omega^{\h1 \t4} (y) &= {1\over 2} \left( \Omega^{\t1 \t4}(y) + \Omega^{\h1\h4} (y) \right)
				   = {1\over 2} \left(- \Omega^{\t2\t3}(1/y) - \Omega^{\h2\h3}(1/y)+\Omega^{\t2\t3}(0) \right)
				   -1  \cr	
\Omega^{\h2\h3}(y)=
\Omega^{\t2 \h3} (y) &= {1\over 2} \left( \Omega^{\t2 \t3}(y) + \Omega^{\h2\h3} (y) \right)	 \,.
\end{aligned}
\eeq

%%%%%%%%%%%%%%%%%%%%%%%%%%%%%%%%%%%%%
%%%%%%%%%%%%%%%%%%%%%%%%%%%%%%%%%%%%%

\appendix{Simple $\mathfrak{su}(2)$}

The simplest and most-well studied example is that of a circular string solution in $S^3 \times \mathbb{R}$. The purpose of this section is not to simply trod on well-explored territory but to make a point in simplifying the computation of fluctuation energies and thus demonstrating the efficiency of our method. 

The circular string in $S^3 \times \mathbb{R}$ is a solution to the $SU(2)$ principal chiral model
\beq
S= \int d\sigma_0 d\sigma_1 \left( \hbox{Tr} J_\alpha^2  + (\partial_\alpha X^0)^2 \right)\,,
\eeq
defined on a Riemann surface $\Sigma$, which determines a map $g: \Sigma \rightarrow SU(2)$ and $X_0: \Sigma \rightarrow \mathbb{R}$. The invariant currents are constructed by $J_{\alpha} =  g^{-1} \partial_{\alpha} g$. The equations of motion supplemented by the Virasoro constraint
\beq
{1\over 2}\hbox{Tr} (J_{\pm}^2) = - \kappa^2
\eeq
ensure that this is a solution to the classical string sigma-model.
Due to classical integrability the equations of motion are equivalent to zero-curvature equations for
\beq
A_{\pm} (x) = {J_{\pm} \over 1\mp x} \,,
\eeq
where $x\in\mathbb{C}$ is the so-called spectral parameter. The monodromy of this flat connection around the $\sigma_1$ spatial circle of the worldsheet is
\beq
\Omega (x) =  P \exp \left[ {1\over 2}\int d\sigma_1 \left(   {J_+ \over 1-x} - {J_- \over 1+x} \right) \right] \,.
\eeq
Unimodularity allows us to diagonalize the monodromy matrix to
\beq
\Omega (x) \equiv \hbox{diag} \left( e^{i p(x)} , e^{- i p(x)}\right) \,.
\eeq
We parametrize $S^3: |Z_1|^2 + |Z_2|^2 =1$ and
\beq
g = \left( \begin{aligned} Z_1 &\quad  Z_2 \cr - \bar{Z_2} &\quad  \bar{Z}_1 \end{aligned} \right) \,.
\eeq
%{\bf Classical solution: explicit form: }

The quasi-momenta for the circule string in $S^3 \times \mathbb{R}$ depend  on the following parameters of the solution, which are the spin $J$ and winding $m$ repackaged as $\mathcal{J} = J/\sqrt{\lambda}$, $\kappa = \sqrt{\mathcal{J}^2 + m^2}$.
The classical energy is
\beq\label{Esu2}
\mathcal{E} = {E \over \sqrt{\lambda}} = \sqrt{\mathcal{J}^2 + m^2} \,.
\eeq
The classical solution is determined by
\beqa
p_{\hat{1}} = p_{\hat{2}} = - p_{\hat{3}} = - p_{\hat{4}}
             & =& + {2 \pi  x \over x^2-1} \kappa\cr
p_{\tilde{1}}& =& + {2 \pi x\over x^2-1} \sqrt{\mathcal{J}^2 + {m^2 \over x^2}} \cr
p_{\tilde{2}}& =& + {2 \pi x\over x^2-1} \sqrt{\mathcal{J}^2 + m^2 x^2} - 2 \pi m  \cr
p_{\tilde{3}}& =& - {2 \pi x\over x^2-1} \sqrt{\mathcal{J}^2 + m^2 x^2}  + 2 \pi m\cr
p_{\tilde{4}}& =& -{2 \pi x\over x^2-1} \sqrt{\mathcal{J}^2 + {m^2 \over x^2}} \,.
\eeqa
The fluctuations were first determined from the sigma-model point of view in  \cite{Frolov:2003tu, Frolov:2004bh}, the exact expansion in terms of $1/\mathcal{J}$ as provided in \cite{SchaferNameki:2006gk} and a derivation of the fluctuation frequencies using the algebraic curve was done in \cite{Gromov:2007aq}. Here we will argue that we only need two frequencies, namely the so-called "internal fluctuations" within the $S^3$ and one $AdS$-fluctuation.

The off-shell frequencies in the $(\t 2, \t 3)$ and $(\h 2, \h3)$ directions are
\beq
\begin{aligned}\label{InputOmegas}
\Omega^{\t 2 \t 3} (y) & = {2 m+ n_{\t2\t3} \over  \kappa y}
         			  = {2 m +{p_{\t 2} -p_{\t 3} \over 2 \pi} \over  \kappa y}
			          = \frac{2 \sqrt{m^2 y^2+\mathcal{J}^2}}{\left(y^2-1\right)
                              \sqrt{m^2+\mathcal{J}^2}}			  \cr
\Omega^{\h 2 \h 3} (y) & = {2 \over y^2 -1} \,.
\end{aligned}			
\eeq
This will be our only input. 
%{\bf Relate with Vicedo's $q(x)$, seems to be the same as off-shell frequency!}.
We will now demonstrate that the remaining $\mathfrak{su}(2)$ frequencies can be obtained by (\ref{sec:QTools}).

The AdS-frequencies are all obtained by generalizations of (\ref{OmegaMapAdS}) 
\beq
\begin{aligned}
\Omega^{\h1 \h4} (y) &= -\Omega^{\h2\h3} (1/y) -2
 				  = {2 \over y^2-1} \cr
\Omega^{\h2 \h 4} (y) & = {1\over 2} \left(\Omega^{\h2 \h3} + \Omega^{\h1 \h 4} \right)
				  = {2 \over y^2 -1} \cr
\Omega^{\h1\h3} (y) & = 	-\Omega^{\h2\h4} (1/y) -2
 				  = {2 \over y^2-1}	\,.		
\end{aligned}
\eeq
Thus showing the expected agreement of all AdS fluctuation energies.

Let us move to the less trivial $S^5$ fluctuations.
From (\ref{FreqTable}) we know
\beq
\Omega^{\t 1\t4} (y) = -\Omega^{\t 2\t3} (1/y)+\Omega^{\t2\t3}(0)   \,.
\eeq
Applied to (\ref{InputOmegas}) we get
\beq
\Omega^{\t 1\t4} (y)
=
\frac{2 \left(-\mathcal{J} y^2+y \sqrt{m^2+y^2 \mathcal{J}^2}
   +\mathcal{J}\right)}{\left(y^2-1\right) \sqrt{m^2+\mathcal{J}^2}}
= {n_{\t1\t4}  y - 2 \mathcal{J} \over \kappa} \,,
\eeq
by recalling that $n_{\t1\t4} = {p_{\t 1}(y)- p_{\t 4} \over 2\pi}$.
The remaining frequencies are obtained by linear combination and inversion
\beq
\begin{aligned}
\Omega^{\t1\t3} (y) &= {1\over 2} \left(\Omega^{\t 1\t4} + \Omega^{\t2\t3} \right)
                                 =   \frac{-\mathcal{J} y^2+\sqrt{m^2+y^2 \mathcal{J}^2}
   y+\mathcal{J}+\sqrt{m^2 y^2+\mathcal{J}^2}}{\left(y^2-1\right)
   \sqrt{m^2+\mathcal{J}^2}} \cr
                                & =
                                {y ( m+n_{\t1\t3} ) - \mathcal{J} - \sqrt{m^2 y^2 + \mathcal{J}^2}\over \kappa} \cr
\Omega^{\t2\t4} (y) & = -\Omega^{\t1\t3} (1/y) - 2  {\partial \mathcal{E} \over \partial \mathcal{J}}  =  \frac{-\mathcal{J} y^2+\sqrt{m^2+y^2 \mathcal{J}^2}
   y+\mathcal{J}+\sqrt{m^2 y^2+\mathcal{J}^2}}{\left(y^2-1\right)
   \sqrt{m^2+\mathcal{J}^2}} \cr
                                 &={y ( m+n_{\t2\t4} ) - \mathcal{J} - \sqrt{m^2 y^2 + \mathcal{J}^2}\over \kappa}   \,.
\end{aligned}
\eeq
Finally we compute the fermion frequencies, which are simply linear combinations
\beq
\begin{aligned}
\Omega^{\h1 \t 4} (y) &= \Omega^{\t1\t4} (y) + \Omega^{\h1 \t1} (y)
             			= \Omega^{\t1\t4} (y) +{1\over 2} \left( \Omega^{\h1 \h4}(y)-\Omega^{\t1\t4}(y) \right)
				= {1\over 2} \left( \Omega^{\h1 \h4}(y)+ \Omega^{\t1\t4}(y) \right)\cr
				&= \frac{-\mathcal{J} y^2+\sqrt{m^2+y^2 \mathcal{J}^2}
   y+\mathcal{J}+\sqrt{m^2+\mathcal{J}^2}}{\left(y^2-1\right) \sqrt{m^2+\mathcal{J}^2}} \cr
                                &= {n_{\h1 \t4} y - \mathcal{J} - \kappa \over \kappa}\cr
\Omega^{\t1 \h3} (y) &= \Omega^{\t1\t4} (y) + \Omega^{\t4 \h3} (y)
             			= \Omega^{\t1\t4} (y) +{1\over 2} \left( \Omega^{\h2 \h3}(y)-\Omega^{\t1\t4}(y) \right)
				= {1\over 2} \left( \Omega^{\h2 \h3}(y)+ \Omega^{\t1\t4}(y) \right)\cr
				&=  {n_{\h1 \t4} y - \mathcal{J} - \kappa \over \kappa}\,.
\end{aligned}
\eeq
Similarly one can check the other fermionic frequencies
\beq
\Omega_{\h1 \t3} (y) = {1\over 2} \left( \Omega_{\t2\t3} (y) + \Omega_{\h1 \h4}(y)\right)
				= {m+  n_{\h1\t3} \over y \kappa}\,.
\eeq
The complete 1-loop energy shift is obtained by
\beq
\delta E = {1\over 2}\sum_{n \in \mathbb{Z}} \sum_{(ij)} (-1)^{F_{ij}} \Omega^{i j} (x_n^{ij}) \,,
\eeq
where $\Omega^{ij}(x_n^{ij})$ are of course now the on-shell frequencies, obtained by evaluating the off-shell frequencies at the position of the poles $x_{n}^{ij}$ determined in (\ref{PolePosition}). This is in complete agreement with \cite{Frolov:2003tu, Frolov:2004bh, Gromov:2007aq}.

%%%%%%%%%%%%%%%%%%%%%%%%%%%%%%%%%%%%%
%%%%%%%%%%%%%%%%%%%%%%%%%%%%%%%%%%%%%
%%%%%%%%%%%%%%%%%%%%%%%%%%%%%%%%%%%%%

\appendix{Details of the two-cut solutions}
\label{sec:appendixtwocut}

In this appendix we summarize some details for the computation of the frequencies of the general two-cut solution in section \ref{sec:GeneralTwoCut}. The terms appearing in the shift of the quasimomenta (\ref{PShift}) are
\beq
\begin{aligned}
\delta \alpha_+ &= \frac{2 \pi  \delta \Delta  z^2+(z-1) N_h \alpha (z)}{2 z^2} \cr
\delta \alpha_- &= \frac{2 \pi  z^2 \delta \Delta -(z+1) N_h \alpha (z)}{2 z^2}\cr
A & = \frac{4 \pi  (f(-1)-f(1)) N_t}{f(-1)+f(1)}-\frac{2 f(-1) f(1) N_h
   \alpha (z)}{z (f(-1)+f(1))} -\frac{((y+1) f(-1)+(y-1) f(1)) f(y) N_t \alpha (y)}{y^2
   (f(-1)+f(1))}
\,.
\end{aligned}
\eeq

\appendix{Details of the one-loop shift computation}
\label{appendix1loop}
In this appendix we collect some intermediate formulas related to the computations of section \ref{sec5}. 

\subsection{Extra poles}
Solving $\exp(i\t p_1-i\t p_3)=1$ we get
\beq
\ba
\epsilon_2=
&\delta^2\frac{X_-(X_+^2-1)}{16 X_+(X_+-X_-)(X_+ X_--1)} \cr
&+\frac{\delta^4}{256}\(\frac{(X_+^2-1)(X_- X_+^3-2 X_+^2+X_- X_+-{X_-}^2+1){X_-}^2}{X_+^2(X_+-X_-)^3(X_+ X_--1)^3}  \right.\cr
&\left.\qquad \qquad \qquad +i\frac{\Delta}{g}\frac{(X_+^2+1){X_-}^2}{ X_+^2(X_+-X_-)^2(X_+ X_--1)^2}\) 
+ \mathcal{O} (\delta^5) \,.
\ea
\eeq
while from $\exp(i\h p_1-i\t p_3)=1$ we get
\beq
\epsilon_3=\delta^2 \frac{e^{i\tau}}{16(X_+-X_-)}
		+\frac{\delta^4}{256} \(\frac{e^{2i\tau}}{(X_+-X_-)^3}+\frac{i\Delta}{g}\frac{(X_+^2+1)e^{2i\tau}}{(X_+^2-1)^2(X_+-X_-)^2}\) + + \mathcal{O} (\delta^5)\,.
\eeq

\subsection{Unphysical fluctuations}
\beq
c=-\frac{2}{X_-X_++1}+\(\frac{(1-X_+X_-)}{4(X_+X_-+1)^2(X_+^2-1)}
\delta+c.c.\) 
\eeq
and
\beq
x^{S_u}=0\,,\qquad  
x^{F_u}=\frac{X_+X_-^{1/2}-X_-X_+^{1/2}}{X_-^{1/2}-X_+^{1/2}} \,.
\eeq
The weighted sum of these contributions then becomes
\beq
\frac{2\Omega_{S_u}(0)-4\Omega_{F_u}(x^{F_u})}{2}=
\frac{2}{X_-X_++1}-\(\frac{X_+^{1/2}X_-^{3/2}-2X_+X_-+X_+^{-1/2}X_-^{1/2}}{4(X_+^2-1)(X_+X_-+1)^2}\delta^2+c.c.\) \,.
\eeq

%%%%%%%%%%%%%%%%%%%%%%%%%%%%%%%%%%%%%
%%%%%%%%%%%%%%%%%%%%%%%%%%%%%%%%%%%%%
%%%%%%%%%%%%%%%%%%%%%%%%%%%%%%%%%%%%%
%\newpage
%%%%%%%%%%%%%%%%%%%%%%%%%%%%%%%%%%%%%
%%%%%%%%%%%%%%%%%%%%%%%%%%%%%%%%%%%%%
%%%%%%%%%%%%%%%%%%%%%%%%%%%%%%%%%%%%%
\bibliographystyle{JHEP}
\renewcommand{\refname}{Bibliography}
\addcontentsline{toc}{section}{Bibliography}
%\bibliography{GMbib}

\providecommand{\href}[2]{#2}\begingroup\raggedright\endgroup

%%%%%%%%%%%%%%%%%%%%%%%%%%%%%%%%%%%%%%%%%%%%%%%%%%%%%%%
%%%%%%%%%%%%%%%%%%%%%%%%%%%%%%%%%%%%%%%%%%%%%%%%%%%%%%%

\end{document}